\documentclass[%
 reprint,
 amsmath,amssymb,
 aps, prd,preprintnumbers, superscriptaddress, nofootinbib,longbibliography
]{revtex4-1}

\usepackage{graphicx}	
\usepackage{amsmath}	
\usepackage{amssymb}	

\usepackage[english]{babel}
\usepackage{graphicx}
\usepackage{latexsym}
\usepackage{floatflt}
\usepackage{float}
\usepackage{units}
\usepackage{verbatim}
\usepackage[colorlinks=true,citecolor=blue,linkcolor=blue]{hyperref}
\usepackage{amsmath}
\usepackage{gensymb}
\usepackage{mathdots}
\usepackage{color}
\usepackage{xcolor}
\usepackage{mathtools}
\usepackage{lipsum}
\usepackage{orcidlink}

\usepackage{dcolumn}
\usepackage{bm}

\newcommand{\id}{{\rm d}}
\newcommand{\prk}{{\cal P}_{\cal R}}

\begin{document}

\preprint{MIT-CTP-LI/6126}

\title{Little Red Dots from Primordial Black Holes:\\Non-Gaussianity and Accretion Evade $\mu$-Distortion Constraints}

\author{Shyam Balaji\,\orcidlink{0000-0002-5364-2109}}
\affiliation{Physics Department, King’s College London, Strand, London, WC2R 2LS, United Kingdom}
\affiliation{New York University Abu Dhabi, PO Box 129188, Saadiyat Island, Abu Dhabi, UAE}

\author{Bryce Cyr\,\orcidlink{0000-0002-6569-3093}}
\affiliation{Center for Theoretical Physics - A Leinweber Institute, Massachusetts Institute of Technology, Cambridge, MA 02139, USA}

\author{Sarah R.~Geller\,\orcidlink{0000-0003-4126-4662}}
\email[Corresponding Author: ]{sarah.geller@yale.edu}
\affiliation{Department of Astronomy, Yale University, Kline Tower, 266 Whitney Avenue, New Haven, CT 06511, USA}

\author{David I.~Kaiser\,\orcidlink{0000-0002-5054-6744}}
\affiliation{Center for Theoretical Physics - A Leinweber Institute, Massachusetts Institute of Technology, Cambridge, MA 02139, USA}

\author{Dario L.~Lorenzoni\,\orcidlink{0000-0001-9832-8183}}
\affiliation{Department of Physics \& Astronomy, University of Manitoba, Winnipeg, MB R3T 2N2, Canada}
\affiliation{Department of Physics, University of Winnipeg, Winnipeg, MB R3B 2E9, Canada}

\author{Evan McDonough\,\orcidlink{0000-0002-2130-3903}}
\affiliation{Department of Physics, University of Winnipeg, Winnipeg, MB R3B 2E9, Canada}

\author{Wenzer Qin\,\orcidlink{0000-0001-7849-6585}}
\affiliation{Center for Cosmology and Particle Physics, Department of Physics, New York University, New York, NY 10003, USA}

\begin{abstract}
    Recent JWST observations of ``Little Red Dots'' (LRDs) at $z \gtrsim 4$ suggest the existence of early supermassive black holes (SMBHs) with masses $\gtrsim 10^6\,M_\odot$. Their rapid emergence challenges conventional seeding and growth scenarios based on accretion and mergers of Population III stellar remnants, motivating renewed interest in primordial black holes (PBHs) formed from primordial perturbations. Explaining such massive seeds requires power spectrum enhancements that are constrained by cosmic microwave background (CMB) spectral distortions. In this work we perform the first comprehensive Bayesian exploration of non-Gaussian primordial perturbations capable of producing SMBH seeds while remaining consistent with CMB spectral distortion constraints, and examine the role of accretion history in constraining these models. We perform a suite of Markov chain Monte Carlo analyses of power spectrum enhancements (lognormal and cut-power-law) with varying local non-Gaussianity and accretion histories, satisfying the \textit{COBE/FIRAS} $\mu$-distortion constraint with and without the requirement that the model produces SMBH seeds. 
    We find that both models can seed LRDs and evade $\mu$-distortion bounds for $|f_{\rm NL}|= {\cal O}(1)$ and peak amplitude $A= {\cal O} (10^{-3})$, when standard Eddington-limited accretion is also incorporated. Interestingly, the results exhibit a preference for both positive and negative values of $f_{\rm NL}$. The preference for negative $f_{\rm NL}$ is strongly dependent on the accretion history, and is removed if the black holes have a small duty cycle, $f_{\rm duty}\lesssim 0.1$. These results demonstrate that the JWST high-$z$ LRD population can be seeded by primordial perturbations while remaining compatible with CMB data, and highlight the importance of next-generation CMB spectral distortion missions to test this hypothesis.
\end{abstract}

\maketitle

\section{\label{sec:level1}Introduction}

Recent James Webb Space Telescope (JWST) observations have revealed unexpectedly luminous quasars and compact ``Little Red Dots'' (LRDs) at redshifts $z \gtrsim 4$~\cite{2023ApJ...954L...4K,2023Natur.616..266L,2023ApJ...959...39H,2024A&A...691A.145M,2024ApJ...963..129M}, corresponding to supermassive black holes (SMBHs) with inferred masses of $\mathcal{O}(10^6$--$10^9)\,M_\odot$~\cite{Volonteri:2021sfo,Dayal:2024zwq,Hai-LongHuang:2024vvz,Hai-LongHuang:2024gtx,Huang:2024aog}. 
Such rapid assembly is in tension with the idea that SMBHs formed as remnants of Population~III stars~\cite{Madau:2001sc,Li:2006ti,Tanaka:2008bv}, since such light seeds would require sustained super-Eddington accretion to grow sufficiently large~\cite{2021ApJ...907L...1W,Bogdan:2023ilu}, prompting the community to instead consider scenarios such as direct collapse of pristine gas clouds forming massive black holes with characteristic masses of $\mathcal{O}(10^{4-6})\,M_\odot$~\cite{Bromm:2002hb,Koushiappas:2003zn,Begelman:2006db,Lodato:2006hw}, and runaway stellar mergers in dense clusters.
While these mechanisms may contribute, whether they can convincingly explain the full observed abundance of overmassive SMBHs at such early epochs remains unclear. This has revived interest in a primordial origin of LRDs in which black holes form directly from the collapse of large curvature perturbations in the early Universe.

Primordial black holes (PBHs)~\cite{Zeldovich:1967lct,Hawking:1971ei,Carr:1974nx} could have formed when sufficiently overdense regions of space re-entered the horizon during radiation domination, collapsing under their own gravitational attraction. 
Such overdensities would be generated by an early phase of cosmic inflation, if the amplitude of the primordial perturbations exceeded a critical threshold.
PBHs can span a wide mass range, from sublunar to supermassive scales, and have long been studied in connection with dark matter~\cite{Meszaros:1975ef,Carr:2020gox} and as potential SMBH seeds~\cite{Nakama:2016kfq,Hooper:2023nnl}. (For reviews, see Refs.~\cite{Khlopov:2008qy,Carr:2020xqk,greenPrimordialBlackHoles2021,Escriva:2022duf}.)
Unlike astrophysical scenarios, PBHs link the origin of SMBHs directly to the primordial power spectrum, providing a rare bridge between cosmological initial conditions and galactic-scale observations. 
However, this connection comes with tight observational constraints.

Small-scale power inevitably leads to the formation of acoustic waves in the photon-baryon plasma, which then dissipate through diffusion (Silk) damping, heating the plasma and generating cosmic microwave background (CMB) spectral distortions \cite{Sunyaev:1970er,Chluba:2012we,Chluba:2015bqa,Fixsen:1996nj}. A $\mu$-type distortion is sourced for modes that re-enter the horizon early and pass the damping scale at redshifts $5\times10^4 \lesssim z \lesssim 2\times10^6$, corresponding to comoving wavenumbers $50 \lesssim k/{\rm Mpc}^{-1} \lesssim 10^4$, precisely the scales relevant for forming PBHs massive enough to seed SMBHs. 
The \textit{COBE/FIRAS} upper limit, $\mu < 4.7\times10^{-5}$~\cite{Fixsen:1996nj,Bianchini:2022dqh}, implies $\mathcal{P}_{\cal R} (k) \lesssim 10^{-4}$ on these scales, stringently constraining PBHs with initial masses $\gtrsim 10^4\,M_\odot$ for Gaussian perturbations~\cite{Kohri:2014lza,Chluba:2015bqa}. 
Thus, $\mu$-distortions currently represent the most stringent limits on any inflationary scenario capable of producing supermassive PBHs by the immediate collapse of primordial density perturbations upon horizon re-entry, under the assumption of Gaussian perturbations.

Recently, an analysis of PBH-based pathways to LRDs concluded that the direct formation of sufficiently massive PBH seeds is strongly excluded by CMB spectral distortion constraints under standard assumptions, while identifying sufficiently large primordial non-Gaussianity as a possible loophole~\cite{DeLuca:2025nao,Nakama:2017xvq,Unal:2020mts,Hooper:2023nnl,Byrnes:2024vjt}. 
Hence, one possible way to evade these constraints is to invoke scale-dependent non-Gaussianity in the primordial curvature perturbations. 

Since PBHs form from the exponentially suppressed tail of the probability distribution of curvature perturbations, even small positive skewness can enhance their formation rate while leaving the average fluctuation amplitude, and therefore the induced $\mu$-distortion, relatively unchanged~\cite{Byrnes:2012yx,Young:2013oia,Franciolini:2018vbk}. 
This motivated several studies exploring whether large local-type non-Gaussianity could rescue the PBH explanation for SMBH seeds~\cite{Nakama:2017xvq,Unal:2020mts,Gow:2022jfb,Hooper:2023nnl}. 
However, a quantitative calculation of non-Gaussian $\mu$-distortion constraints by Ref.~\cite{Byrnes:2024vjt} showed that even in the extreme non-Gaussian limit, the spectral distortion bound on $\delta$-function enhancements to the primordial power spectrum is not significantly relaxed, underscoring the need to explore more general scenarios.

Interestingly, the effect of negative local non-Gaussianity on PBH formation is not monotonic. Moderately negative values of $f_{\rm NL}$ suppress the PBH abundance, but for realistic finite-width power spectra, the PBH abundance can increase again at large negative $f_{\rm NL}$~\cite{Franciolini:2023pbf}. As we show below, this behavior leads to a second, negative-$f_{\rm NL}$ region compatible with the required SMBH-seed abundance.

Complementary efforts from Ref.~\cite{Wang:2025hbw} incorporated non-Gaussianity into PBH abundance calculations using the compaction-function formalism and showed that broad power spectra, such as lognormal or broken-power-law shapes expected in realistic inflationary models, can significantly alter the allowable region for SMBH seeds by $\mu$-distortion bounds relative to $\delta$-function peaked spectra. However, those authors did not consider the nonlinearities between the Gaussian curvature perturbation and the density contrast arising at horizon crossing, as in Ref.~\cite{Musco:2020jjb}, which can significantly affect the resulting bounds. For simplicity, Ref.~\cite{Wang:2025hbw} also fixed various relations in the compaction computation (e.g., $r_{m} = 1/k$), when in general these proportionalities depend on the exact shape of the power spectrum. 

In addition, Ref.~\cite{Dave2026} considered non-perturbative probability distribution tails while treating the $\mu$-distortion primarily through its dependence on the total variance. Our analysis is complementary: we explicitly include the non-Gaussian loop correction to the $\mu$-distortion, compute the PBH abundance using the compaction function formalism, and jointly infer the allowed power spectrum and local $f_{\rm NL}$ parameter space.

In this work, we unify and extend these directions. 
We perform the first Markov chain Monte Carlo (MCMC) exploration of the parameter space in which non-Gaussianity, power spectrum amplitude, and spectral width are treated jointly, using the full non-Gaussian $\mu$-distortion formalism applied to broad-band spectra, implementing nonlinearities between the curvature perturbation and density contrast. Crucially, we also incorporate accretion following PBH formation. For PBHs that form with masses $M > M_\odot$, accretion should be reasonably efficient \cite{Rice:2017avg,DeLuca:2020fpg,DeLuca:2021pls,DeLuca:2023bcr}, so it is important to consider its effects when comparing with JWST observations at redshifts $z \simeq 7$, long after the PBH formation times in this scenario.

By considering lognormal and cut-power-law templates for the curvature power spectrum, we quantify the level of non-Gaussianity required to produce primordial SMBH seeds while satisfying current $\mu$-distortion limits. 
The resulting posterior distributions delineate the viable region of early-universe parameter space that can explain high-redshift SMBHs and yield exclusion contours for non-Gaussian PBH scenarios based solely on spectral distortion data. 
Our results are broadly applicable to inflationary model-building and will be fully testable by proposed spectral distortion missions, such as the FOSSIL \cite{FOSSIL_EXP}, PIXIE \citep{Kogut2011PIXIE, Kogut2016SPIE, KogutPIXIE2024}, or SPECTER \cite{SPECTER2024} designs.

This paper is organized as follows. 
In Sec.~\ref{sec:level2} we summarize the formalism for PBH abundance, while in Sec. \ref{sec:level3} we outline the $\mu$-distortion calculation, including the effects of non-Gaussian curvature perturbations. Section~\ref{sec:methods} summarizes our methodology, and
Sec.~\ref{sec:results} showcases the MCMC results for multiple power spectrum templates and quantifies the allowed parameter space consistent with both LRD and spectral distortion constraints. We conclude in Sec.~\ref{sec:level5} with implications for inflationary model-building and prospects for future observational tests. Throughout this paper, we work in natural units in which $\hbar = c = k_{\rm B} = 1$.

\section{\label{sec:level2}The Abundance of Supermassive Seeds}

To build intuition, let us first calculate the PBH abundance with a simplified approach, in which we assume the formation of a black hole is inevitable once the smoothed curvature perturbation is larger than a critical threshold value, $\mathcal{R}_{\rm c}$. (See, e.g., Ref.~\cite{Carr:2020xqk}.) The initial mass fraction can then be calculated by integrating the probability distribution function of $\mathcal{R}$ from $\mathcal{R}_{\rm c}$ to $\infty$, 
%
\begin{equation}
    \beta =\int_{\mathcal{R}_{\rm c}}^{\infty} {\rm d} \mathcal{R} \, P(\mathcal{R}) \frac{M_{\rm PBH}}{M_{\rm H}}\label{simplebeta}.
\end{equation}
%
Here $\beta$ denotes the probability for a primordial perturbation to exceed the collapse threshold, which in this simplified treatment is identified with the initial PBH formation fraction, $P(\mathcal{R})$ is the one-point probability distribution of the smoothed curvature perturbation, and $\mathcal{R}_{\rm c}$ is the corresponding collapse threshold. We also introduce $M_{\rm H}$, the horizon mass at the time $t_i$ when a given overdensity (labeled by its comoving wavenumber $k$) re-enters the Hubble radius~\cite{Nakama:2016gzw},
%
\begin{equation}\label{eq:horizon_mass}
    M_{\rm H}(k) \simeq 17 \left(\frac{g_*}{10.75}\right)^{-1/6} \left(\frac{k}{10^6 \, \mathrm{Mpc}^{-1}}\right)^{-2} M_\odot ,
\end{equation}
%
where $g_*$ counts the number of effectively massless degrees of freedom in the particle spectrum. The typical PBH mass at formation $\bar{M}_{\rm PBH}$ (near the peak of the mass distribution) is related to the horizon mass by 
%
\begin{equation}\label{eq:Mpbh-bar}
    \bar{M}_{\rm PBH} (t_i) =\eta\, M_{\rm H} (t_i) ,
\end{equation}
%
with the efficiency factor commonly taken to be $\eta \simeq 0.2$ \cite{Carr:1975qj,Carr:2020xqk}. If the curvature perturbations are Gaussian distributed with a variance $\sigma^2$, then Eq.~\eqref{simplebeta} reduces to \cite{Carr:2020xqk}
%
\begin{equation}
    \beta \propto \text{erfc}\left(\frac{\mathcal{R}_{c}}{\sqrt{2}\sigma}\right).
\end{equation}
%
PBHs are oft-quoted in terms of their relative contribution to the dark matter density today, 
%
\begin{equation}
f_\mathrm{PBH} \equiv \left. \frac{\Omega_\mathrm{PBH}}{\Omega_\mathrm{DM}} \right|_\mathrm{today} ,
\label{fpbhdef}
\end{equation}
%
where $\Omega_{\rm PBH}$ and $\Omega_{\rm DM}$ denote the relative energy density fraction in PBHs and dark matter, respectively.

A common approach to obtain the value of $f_{\text{PBH}}(M)$ today is to scale $\beta$ in Eq.~\eqref{simplebeta} by a factor of $\sqrt{M_{\rm eq}/M_{\rm H}}$ to account for the dilution of radiation and matter during the radiation dominated epoch, where $M_{\rm eq}$ is the horizon mass scale at the epoch of matter-radiation equality, which is approximately $2.8\times 10^{17}M_{\odot}$. This redshifting factor gives the PBH abundance in the absence of subsequent mass growth. For comparison with the LRD population, we separately evolve the primordial seed masses through the accretion prescription described in Sec.~\ref{sec:level2.2}, terminating the evolution at the characteristic LRD redshift $z\simeq7$. The scaling dependence on $M_{\rm eq}$ follows from the observation that PBHs dilute as $a^{-3}(t)$ whereas radiation dilutes as $a^{-4}(t)$, such that the PBH-to-radiation energy density fraction increases during the era of radiation domination. 

However, this approach to calculating  $f_{\rm PBH}$ misses the correlations between perturbations at different scales. It also (incorrectly) assumes that the curvature threshold for precipitating a collapse is constant across all scales. A more rigorous approach uses the so-called \textit{compaction function formalism} \cite{Shibata:1999zs,Musco:2008hv,Musco:2020jjb,Young:2024jsu}, which improves on the simple approximation by recognizing that the collapse criterion is dependent on the spatial scale. 

Using this formalism, PBHs form when the compaction function $\mathcal{C}(r, t)$ associated with an overdensity exceeds a threshold $\mathcal{C}_\mathrm{th}$ upon horizon re-entry. The compaction function is defined as \cite{Musco:2008hv,Musco:2020jjb}:
%
\begin{equation} \label{eq:full_compaction}
\mathcal{C}(r, t) \equiv 2 \frac{G \, \delta M(r, t)}{R(r, t)},
\end{equation}
%
where $\delta M(r,t)$ is the mass excess enclosed within the areal radius $R(r,t)$, and $G$ is the Newton's constant. The resulting PBH mass follows a critical collapse scaling law \cite{Choptuik:1992jv,Evans:1994pj,Gundlach:2007gc}
%
\begin{equation}
M_\mathrm{PBH} = \mathcal{K} M_{\rm H} (\mathcal{C} - \mathcal{C}_\mathrm{th})^\gamma,
 \end{equation}
%
where $\mathcal{K} = 4.4$ and $\gamma = 0.38$ (for collapse within a radiation-dominated fluid \cite{Musco:2020jjb}). The curvature perturbation power spectrum is represented generically as $\mathcal{P}_\mathcal{R}(k)$.

In this setup, the full curvature perturbation can be written as a sum of Gaussian and non-Gaussian parts,
%
\begin{equation}
    \mathcal{R}=\mathcal{R}_{\rm G}+\mathcal{R}_{\rm NG} ,
\end{equation}
%
where the non-Gaussian part $\mathcal{R}_{\rm NG}$ depends non-trivially on the Gaussian random field $\mathcal{R}_{\rm G}$. The former can be expanded in a series expansion as $\mathcal{R}_{\rm NG}=\frac{3}{5}f_{\rm NL}\mathcal{R}_{\rm G}^2+\frac{9}{25}g_{\rm NL}\mathcal{R}_{\rm G}^3+\dots$, where the parameters $f_{\rm NL},\,g_{\rm NL}$ capture the amplitudes of (local-type) non-Gaussian corrections to the bispectrum and trispectrum, respectively. In this work we restrict to the case of quadratic local non-Gaussianity, setting higher-order coefficients such as $g_{\rm NL}$ to zero. Thus the non-Gaussian mapping relevant for our numerical analysis is given by
%
\begin{equation}
    \mathcal{R}=\mathcal{R}_{\rm G}+\tilde{f}_{\rm NL}(\mathcal{R}_{\rm G}^{2} - \langle \mathcal{R}_{\rm G}^2 \rangle) ,
    \label{eq:Rform}
\end{equation}
where we define
%
\begin{equation}
    \tilde{f}_{\rm NL} \equiv \frac{3}{5}f_{\rm NL}.
\end{equation}
%
The analyses presented in this work will consider both the case of perturbative non-Gaussianity, where the above is a good approximation to the series expansion, and {\it non-perturbative} non-Gaussianity, where Eq.~\eqref{eq:Rform} is interpreted as a toy model, and not the leading order terms in a series expansion.

\subsection{\label{sec:level2.1}The primordial mass function}

We compute the present-day abundance of PBHs formed from large primordial curvature perturbations during radiation domination using both lognormal and cut-power-law parameterizations for the power spectrum (which we detail below). 

Under the assumption of a spherically symmetric curvature perturbation, one can write the full compaction function $\mathcal{C}$ given in Eq.~\eqref{eq:full_compaction} as \cite{LISACosmologyWorkingGroup:2023njw}
%
\begin{align}
    \mathcal{C} &= \mathcal{C}_l - \frac{1}{4 \Phi} \mathcal{C}_l^2,\\
    \mathcal{C}_l &= \mathcal{C}_{\rm G} \, \frac{{\rm d} \mathcal{R}}{{\rm d} \mathcal{R}_{\rm G}},\\
    \mathcal{C}_{\rm G} &= - 2 \Phi r \, \frac{{\rm d} \mathcal{R}_{\rm G}}{ {\rm d} r}.
\end{align}
%
Here $\mathcal{C}_{l}$ and $\mathcal{C}_{\rm G}$ are referred to as the \textit{linear} and \textit{Gaussian linear} components of the compaction, while during radiation domination the Bardeen potential is simply given by $\Phi = 2/3$. 
We work in comoving slicing and use the long-wavelength approximation, in which the spherically symmetric spatial metric takes the form $a^2(t)e^{2\mathcal R(r)} (dr^2+r^2d\Omega^2)$. Here $r$ denotes the comoving radial coordinate, while $R(t,r)=a(t)e^{\mathcal R(r)}r$ is the corresponding areal radius appearing in Eq.~\eqref{eq:full_compaction}.
The characteristic scale $r_m$ denotes the comoving coordinate at which the compaction function reaches its maximum. We associate this scale with the background horizon-crossing wavenumber through $r_m\simeq k_{\rm H}^{-1}$, noting that the exact horizon-entry condition involves the areal radius, $H(t_{\rm H})R(t_{\rm H},r_m)\simeq1$. $k_\textrm{H}$ is the horizon-crossing wavenumber associated with horizon mass $M_\textrm{H}$; this should not be confused with the peak scale $k_p$ of the primordial power spectrum (see Eqs.~\eqref{lognormalPR} and~\eqref{sharkPR}), since the relation between $k_p r_m$ and the spectral shape is determined self-consistently. Rearranging Eq.~\eqref{eq:horizon_mass} we find 
%
\begin{align}
    r_m \simeq 2.4\times 10^{-7} {\rm Mpc}\left(\frac{g_*}{10.75}\right)^{1/12}\left(\frac{M_\textrm{H}}{M_\odot}\right)^{1/2}.
\end{align}
%
Note that when the curvature perturbation follows purely Gaussian statistics, $\mathcal{C}_l = \mathcal{C}_{\rm G}$. 

Within this formalism, the computation of the initial mass proceeds in a similar way to Eq.~\eqref{simplebeta}
%
\begin{align} \label{eq:beta_fun_comp}
    \beta_{\rm \mathcal{C}} = \int_{\rm \mathcal{D}} {\rm d} \mathcal{C}_{\rm G} \, {\rm d} \mathcal{R}_{\rm G} \, P_{\rm G}(\mathcal{C}_{\rm G}, \mathcal{R}_{\rm G}) \, \frac{M_{\rm PBH}}{M_{\rm H}},
\end{align}
%
with $P_{\rm G}$ being the bivariate Gaussian probability distribution function (PDF) describing the statistics of the small-scale fluctuations. One condition on the integration range $\mathcal{D}$ is determined by the requirement that collapse only proceeds when the compaction exceeds a given threshold ($\mathcal{C} \geq \mathcal{C}_{\rm th}$). The collapse threshold $\mathcal{C}_{\rm th}$ is not taken to be a universal constant. Following Ref.~\cite{Musco:2020jjb}, we determine it from the characteristic perturbation profile associated with the input power spectrum, so that its value depends on the width and shape of $\mathcal{P}_{\mathcal R}(k)$. Given $\beta_{\mathcal{C}}$, it is straightforward but cumbersome to calculate the mass spectrum of PBHs at some later time. As a result, we quote the result and provide details of the computation in Appendix~\ref{sec:levelA0}:
%
\begin{align} \label{eq:mass_spectrum}
    \frac{{\rm d} \Omega_{\rm PBH}}{{\rm d} \log M_{\rm PBH}} = &\frac{\mathcal{K}}{\gamma} \int_{M_{\rm H}^{\rm min}}^\infty \frac{{\rm d} M_{\rm H}}{M_{\rm H}} \left( \frac{M_{\rm eq}}{M_{\rm H}}\right)^{1/2} \left(\frac{M_{\rm PBH}}{\mathcal{K} M_{\rm H}} \right)^{\frac{1+\gamma}{\gamma}} \nonumber \\
    &\times \left[1 - \frac{\mathcal{C}_{\rm th}}{\Phi} - \frac{1}{\Phi} \left(\frac{M_{\rm PBH}}{\mathcal{K} M_{\rm H}} \right)^{1/\gamma} \right]^{-1/2} \nonumber \\
    &\times \int {\rm d} \mathcal{R}_{\rm G} \left( \frac{{\rm d} \mathcal{R}}{{\rm d} \mathcal{R}_{\rm G}} \right)^{-1} P_{\rm G}(\mathcal{C_{\rm G}}[M_{\rm H}], \mathcal{R}_{\rm G}).
\end{align}
%
The minimum horizon mass capable of seeding a PBH is given by $M_{\rm H}^{\rm min} = M_{\rm PBH}/[\mathcal{K} (\Phi - \mathcal{C}_{\rm th})^{\gamma}]$, as detailed in Appendix~\ref{sec:levelA0}. With this, the integrated fractional energy density in PBHs is readily calculated as
%
\begin{align} \label{eq:f_pbh}
    f_{\rm {PBH}} = \frac{1}{\Omega_{\rm DM}} \int \frac{{\rm d} M_{\rm PBH}}{M_{\rm {PBH}}} \frac{{\rm d} \Omega_{\rm PBH}}{{\rm d} \log M_{\rm PBH}}.
\end{align}
%
In practice, we evaluate the integrals in Eqs.~\eqref{eq:mass_spectrum} and \eqref{eq:f_pbh} numerically.

\subsubsection*{Power spectrum templates}

\begin{figure*}
    \centering
    \includegraphics[width=.95\linewidth]{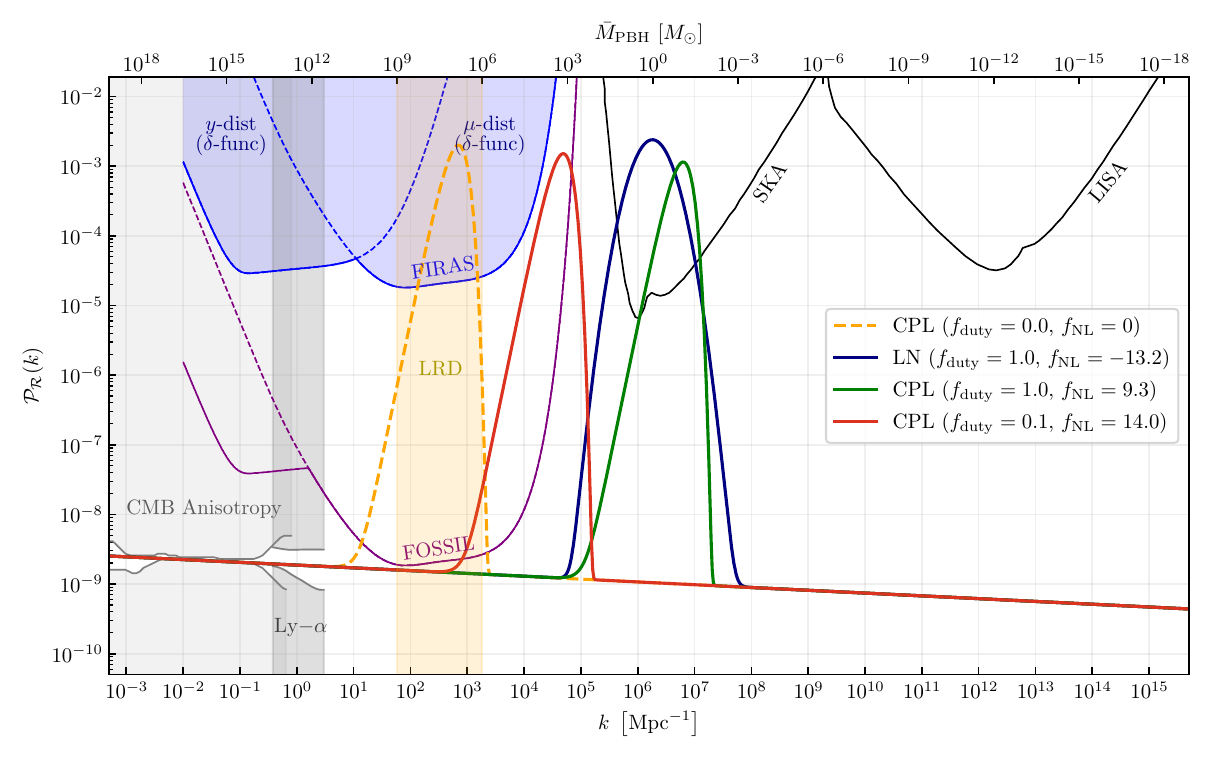}
    \caption{Curvature power spectra ${\cal P}_{\cal R}(k)$ for the lognormal and cut-power-law templates (Eqs.~\eqref{lognormalPR} and~\eqref{sharkPR}) that satisfy current $\mu$-distortion constraints and seed JWST LRDs. 
    We superimpose our spectrum templates onto the baseline $\Lambda$CDM power spectrum, ${\cal P}_{\cal R}^{\Lambda \rm CDM}=A_s(k/k_\star)^{n_s-1}$ \cite{Planck:2015sxf} with the \textit{Planck} best-fit values for $A_s$ and $n_s$ \cite{Planck2018}. 
    For our power spectrum templates, we consider the best-fit parameters given in the right-hand columns of Tables~\ref{tab:ln-results} and~\ref{tab:cpl-results}, and in the left-hand column of Tab.~\ref{tab:cpl-results-fduty}; the navy, green, and red curves have $f_{\rm NL}=\{-13.2,\, 9.3,\, 14.0\}$, respectively.
    The current \textit{COBE/FIRAS}~\cite{Fixsen:1996nj,Bianchini:2022dqh} $\mu$- (and $y$-) distortion constraints are shown in blue, with the forecast sensitivity of FOSSIL~\cite{FOSSIL_EXP, Cyr_SD_whitepaper} shown in purple. We also show the constraints from CMB anisotropies~\cite{Planck:2018jri} and the Lyman-$\alpha$ forest~\cite{2011MNRAS.413.1717B} in gray, and the expected sensitivity of the gravitational wave experiments SKA~\cite{2009IEEEP..97.1482D,Gow:2020bzo} and LISA~\cite{LISACosmologyWorkingGroup:2022jok,Gow:2020bzo} in black. 
    We highlight in orange the SMBH mass range corresponding to the JWST LRDs (the upper horizontal axis indicates the typical PBH mass at formation, Eq.~\eqref{eq:Mpbh-bar}).
    We point out that, without accretion or non-Gaussianity, models that could directly produce the JWST LRDs (dashed orange curve) are ruled out by $\mu$-distortion constraints. For sufficiently efficient accretion ($f_{\rm duty} \simeq 1$), Gaussian fluctuations may satisfy both constraints, whereas reduced accretion efficiency increasingly favors non-Gaussian statistics.
    }
    \label{fig:prk}
\end{figure*}

We consider two benchmark forms for the curvature power spectrum, motivated by inflationary models producing enhanced small-scale fluctuations. We adopt both a lognormal and a cut–power-law template. In the vicinity of the peak, the lognormal spectrum is given by 
%
\begin{equation}
    \label{lognormalPR}
    \mathcal{P}^{\rm LN}_\mathcal{R}(k)
     = {A}\,
     \exp\!\left[-\frac{\ln^2(k/k_p)}{2\Delta^2}\right] ,
\end{equation}
%
where $A$ sets the overall amplitude of the peak, \(k_p\) denotes its characteristic scale and \(\Delta\) controls its width in logarithmic space. 
This form captures spectra peaked around a single inflationary scale and allows the width parameter to encode how many modes participate in collapse.

The cut–power-law spectrum is instead defined as%
\footnote{
    Note that this expression is equivalent to the form of the cut-power-law spectrum often found in literature~\cite{Musco:2020jjb}, $\prk(k)=A_*(k/k_*)^{n_*} e^{-(k/k_*)^2}$, where we have rescaled $A\equiv A_* [n_*/(2e)]^{n_*/2}$ and $k_p\equiv k_*\sqrt{n_*/2}$. With this rescaling, the peak of the power spectrum is located at $k_p$ with amplitude $\prk(k_p)=A$, making explicit that $A$ is the maximum amplitude of the spectrum, occurring at a scale $k_p$, consistent with the case of the lognormal power spectrum. For both spectrum templates, we sample over the amplitude $A$ and the characteristic scale $k_p$.
}
%
\begin{equation}
    \label{sharkPR}
    \mathcal{P}_\mathcal{R}^{{\rm CPL}}(k)
    = {A}
    \left(\frac{k}{k_p}\right)^{n_*}
    \exp\!\left[\frac{n_*}{2}\left(1-\left(\frac{k}{k_p}\right)^{2}\right)\right] ,
\end{equation}
%
where \(n_*\) is the effective tilt and the exponential provides a smooth ultraviolet cutoff. We fix $n_*=4$ as a representative steep-growth benchmark motivated by the approximately $k^4$ rise obtained in single-field inflationary scenarios that generate PBH-producing enhancements of the curvature spectrum \cite{Byrnes:2018txb,Liu:2020oqe}.
The total power spectrum is thus given by the sum of the $\Lambda$CDM contribution and the enhancement, as
%
\begin{align}
    \mathcal{P}_{\mathcal{R}}(k) = \mathcal{P}_{\mathcal{R}}^{\Lambda {\rm CDM}} (k)+ \mathcal{P}_{\mathcal{R}}^{\rm LN/CPL} (k)
\end{align}
%
where the $\Lambda$CDM floor is given by the usual parametrization ${\cal P}_{\cal R}^{\Lambda \rm CDM}=A_s(k/k_{\rm CMB})^{n_s-1}$ \cite{Planck:2015sxf} and we consider the \textit{Planck} best-fit values for $A_s$ and $n_s$ and the pivot scale $k_{\rm CMB}=0.05~{\rm Mpc}^{-1}$~\cite{Planck2018}.

We illustrate these power spectrum templates in Fig.~\ref{fig:prk}, where we overlay the lognormal and cut-power-law power spectra with observational constraints. Parameter values are taken to be best-fit values in the analyses reported later in this work, namely in Tables~\ref{tab:ln-results},~\ref{tab:cpl-results}, and~\ref{tab:cpl-results-fduty}. From this figure, one may appreciate the projected constraining power of the Square Kilometer Array (SKA) ~\cite{2009IEEEP..97.1482D,Gow:2020bzo} and of the proposed CMB spectral distortion satellite FOSSIL \cite{FOSSIL_EXP, Cyr_SD_whitepaper}.
Notably, enhancements of the curvature power spectrum that could directly produce the SMBHs corresponding to JWST LRDs (orange shading) without considering accretion or non-Gaussian perturbations (dashed orange curve) are ruled out by $\mu$-distortion constraints.

\subsection{\label{sec:level2.2}Mergers and acquisitions}
Once the PBHs have formed, their masses continue to evolve due to accretion of surrounding matter and mergers with other PBHs.
Understanding how the PBH mass function changes with time then allows us to test a given model against observations made today.
For this work, we will neglect the effect of mergers, since their impact is heavily model-dependent and we expect them to only change the final mass of a PBH by an $\mathcal{O}(1)$ factor. We model accretion onto the PBH by assuming it accretes at or below the Eddington limit, i.e. its luminosity is given by
%
\begin{equation}
    L_\mathrm{Edd} = \frac{4\pi G M_\mathrm{PBH}\, m_p}{\sigma_T},
    \label{eq:L_eddington}
\end{equation}
%
where $m_p$ is the proton mass and $\sigma_T$ is the Thomson cross-section, and the black hole's growth rate is related to the luminosity by
%
\begin{equation}
    \dot{M}_\mathrm{Edd} = (1-\epsilon)\frac{f_\mathrm{duty} L_\mathrm{Edd}}{\epsilon}.
    \label{eq:eddington}
\end{equation}
%
Here $\epsilon$ is a factor characterizing how efficiently the accreted matter is converted into radiation, which we take to be $\epsilon=0.1$. 
\footnote{
    We adopt the value $\epsilon=0.1$, commonly considered in the literature~\cite{2021ApJ...907L...1W,Bogdan:2023ilu}, to err on the side of caution. 
    In the limit of zero spin, we calculate the efficiency factor to be $\epsilon=0.057$.
    While we do not expect the PBH population studied in this work to be produced with significant amounts of spin, using a lower value of $\epsilon$ would make accretion more efficient, thus allowing PBHs to seed the JWST LRDs with a smaller initial mass.
    In this sense, we consider $\epsilon=0.1$ to be a conservative value.
}
The duty cycle parameter, $f_\mathrm{duty}$, parametrizes the fraction of time during which the black hole is actively accreting. In our analyses we assume $f_\mathrm{duty} = 1$ unless otherwise stated, and perform a separate analysis of the effect of smaller values of $f_{\rm duty}$. Typical values assumed for very active quasars are $0.3 \lesssim f_{\rm duty} \leq 1$~\cite{Tanaka:2008bv,2013ApJ...771..116J,Milosavljevic:2008pi,Pacucci:2021ubg}, although some studies indicate that $f_{\rm duty}$ could potentially be much smaller for primordial seeds~\cite{Dayal:2025aiv}.

While we include constraints from spectral distortions, we will neglect limits based on accretion~\cite{Ali-Haimoud:2016mbv,Serpico:2020ehh,Agius:2024ecw}.
Recent work has reexamined CMB anisotropy constraints on PBH accretion (for Gaussian statistics)~\cite{Agius:2024ecw}, finding that the constraint on $f_\mathrm{PBH}$ at a black hole mass of $M = 10^4 \,M_\odot$ can vary between $10^{-8} - 10^{-5}$ depending on the accretion model, or between $10^{-5} - 10^{-2}$ for $M = 10^2 \,M_\odot$. 
We will be interested in LRD abundances that roughly correspond to $f_\mathrm{PBH} \sim 10^{-7}$ at the redshift of the LRDs~\cite{Unal:2020mts}, and an even smaller value at the time of PBH formation, due to accretion.
Moreover, the accretion bounds listed in the above references do not include non-Gaussianities and must be recomputed in this context. We leave this as a direction for future work.

\begin{figure}[h!]
    \centering
    \includegraphics[width=\linewidth]{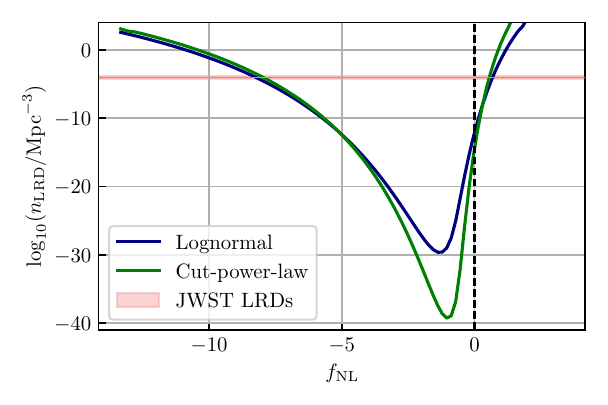}
    \caption{
    Predicted comoving LRD number density as a function of local non-Gaussianity $f_{\rm NL}$, with the remaining power-spectrum parameters held fixed at $\log_{10}{A}=-2.3$ and $\log_{10}(k_p/{\rm Mpc}^{-1})=5.5$. For the lognormal template we take $\Delta=0.5$, while the cut-power-law template has $n_*=4$. The horizontal band denotes the observed LRD abundance, $\log_{10}(n_{\rm LRD}/{\rm Mpc}^{-3})=-4.0\pm0.1$. The two intersections illustrate that both positive- and negative-$f_{\rm NL}$ are capable of reproducing the observed abundance.}
    \label{fig:fNLnLRD}
\end{figure}

\subsection{Little Red Dots}\label{sec:LRD}

JWST observations indicate a substantial population of compact LRDs at $z_{\rm LRD} \simeq 7$. 
Following Ref.~\cite{2024ApJ...968....4P}, we adopt the observed comoving number density%
\footnote{
    We follow the result of Ref.~\cite{2024ApJ...968....4P}, commonly adopted in the literature (see e.g. Refs.~\cite{Zubovas:2026hqc,Ma:2025qna}), to remain conservative. Recent observations seem to indicate a slightly lower abundance of LRDs~\cite{Herrero-Carrion:2026back,Inayoshi:2025isg,Inayoshi:2025critical}; a smaller value of $n_{\rm LRD}$ would make it easier for our PBH seeds to match the LRD population, by requiring less support for the high-mass tail of the probability distribution of non-Gaussian primordial perturbations. It would therefore be possible to accommodate a lower $n_{\rm LRD}$ with smaller $f_{\rm NL}$.
}
%
\begin{equation}
    \log_{10}\!\left(\frac{n_{\rm LRD}}{{\rm Mpc}^{-3}}\right)
    =-4.0\pm0.1 .
\end{equation}
%
For the purposes of this work we interpret this abundance as a target number density of actively accreting massive black holes descended from primordial seeds. This identification is a benchmark assumption: the astrophysical nature of the full LRD population remains uncertain, and not every observed LRD needs to host an accreting massive black hole.

The primordial PBH mass function derived in Sec.~\ref{sec:level2.1} can be converted into a comoving seed number density as
%
\begin{equation}
    \frac{{\rm d}n_{\rm PBH}}{{\rm d}\ln M_i}
    =
    \frac{\rho_{\rm DM,0}}{M_i}
    \frac{{\rm d}f_{\rm PBH}}{{\rm d}\ln M_i},
\end{equation}
%
where $M_i$ denotes the primordial seed mass and $\rho_{\rm DM,0}$ is the present-day mean dark matter density. In the absence of mergers the comoving number of black holes is conserved, while accretion maps each initial mass $M_i$ to a later mass $M_f(M_i,z)$. The predicted LRD abundance can therefore be written schematically as 
%
\begin{equation}
\begin{split}
    n_{\rm model}(z_{\rm LRD})
    =
    \int& {\rm d}\ln M_i\,
    \frac{\rho_{\rm DM,0}}{M_i}
    \frac{{\rm d}f_{\rm PBH}}{{\rm d}\ln M_i} \\
    &\quad\times
    {\cal S}\!\left[M_f(M_i,z_{\rm LRD})\right],
\end{split}
\end{equation}
%
where ${\cal S}$ selects black holes whose accreted masses lie in the LRD mass range. 
Our fiducial calculation adopts the accretion prescription described in Sec.~\ref{sec:level2.2}.
We take a simple prescription for the LRD mass range and simply require that the black hole mass at $z_{\rm LRD} \simeq 7$ be greater than $10^6 \,M_\odot$.
Since we assume that the black hole luminosity is given by the Eddington limit, as in Eq.~\eqref{eq:L_eddington}, this corresponds to objects with luminosity greater than $L \sim 10^{44}$~erg/s~\cite{2024ApJ...964...39G}.
We note that recent work suggests that the intrinsic luminosities of the LRDs may be somewhat smaller than originally determined, which would correspond to smaller LRD masses~\cite{Greene:2025njv}. For a given accretion model, taking a smaller mass/luminosity threshold would allow for smaller initial PBH seeds, allowing our results described later in Section~\ref{sec:results} to shift towards enhancements at larger wavenumbers.

The strong dependence of the predicted LRD abundance on primordial non-Gaussianity is illustrated in Fig.~\ref{fig:fNLnLRD}. Holding the power spectrum parameters fixed, the abundance is extremely small near the Gaussian limit, while positive $f_{\rm NL}$ rapidly enhances PBH formation by increasing the probability of rare over-threshold fluctuations. Importantly, the dependence is not monotonic for negative $f_{\rm NL}$. For sufficiently negative $f_{\rm NL}$, configurations in which $\mathcal{R}_{\rm G}$ and $\mathcal{C}_{\rm G}$ are anti-correlated can again satisfy the compaction threshold, causing the PBH and hence LRD abundance to increase. Consequently, the observed LRD number density can be reproduced for both a positive and a negative $f_{\rm NL}$. This behavior provides a simple physical interpretation of the bimodality found in the MCMC analysis below.

\section{\label{sec:level3}CMB Spectral Distortions from Primordial Perturbations}

Physically, enhancements to the primordial curvature spectrum lead to a larger patch-to-patch variance in the temperature fluctuations at horizon re-entry. The dissipation of these fluctuations through diffusion (Silk \cite{Silk1967}) damping is an inherently non-thermal process, capable of producing either $\mu-$ or $y-$type spectral distortions dependent on the redshift at which the patches mix \cite{Chluba2012a, Chluba:2012we, Khatri2012, Pajer2012, Chluba2013, Chluba2015, Cyr2023}. Roughly speaking, $\mu$-type distortions are sourced by non-thermal energy and entropy injections between $5\times 10^4 \lesssim z \lesssim 2 \times 10^6$, whereas $y$-type distortions are produced at later times. Thus, late-time energy injections from post-recombination sources (e.g. reionization) can produce an irreducible background $y$ signal which makes it difficult to disentangle early-time sources. 

The $\mu$-type distortion, however, is a clean probe\footnote{It is notoriously difficult to produce a $\mu$-type distortion at low ($z \lesssim 10^4$) redshift, though it has recently been pointed out that it may be possible through resonant photon to dark photon conversions \cite{Chluba2024, Arsenadze2024}.} of the primordial Universe, and is primarily sensitive to the power spectrum on scales $50 \, {\rm Mpc}^{-1} \lesssim k \lesssim 10^4 \, {\rm Mpc}^{-1}$. If one assumes that the power spectrum remains Gaussian and nearly scale invariant (following central values inferred by the Planck satellite \cite{Planck2018}) on these small scales, a distortion with amplitude $\mu \simeq 2 \times 10^{-8}$ is generated \citep{Chluba2012a, Cabass2016, Chluba2016}. This is a far cry from the sensitivity of the \textit{COBE/FIRAS} satellite \cite{Fixsen1996}, which set a (recently improved \cite{Bianchini:2022dqh}) bound of $|\mu| \lesssim 4.7 \times 10^{-5}$ ($95\%$ confidence level). This does, however, present a clear science case for next-generation spectral distortion satellites such as FOSSIL \cite{FOSSIL_EXP, Cyr_SD_whitepaper}, PIXIE \citep{Kogut2011PIXIE, Kogut2016SPIE, KogutPIXIE2024}, and SPECTER \cite{SPECTER2024}, all of which have a projected sensitivity capable of detecting the Silk damping distortion at the $4\sigma$ level or above. Excitingly, FOSSIL has recently advanced through a pre-selection process as part of the M8 mission opportunity within the European Space Agency (ESA). CMB spectral distortions have also been identified by ESA as a potential science target as part of their Voyage 2050 space program \cite{Chluba2019}.

The formation of PBHs through conventional means requires a large enhancement to the amplitude of the primordial spectrum.\footnote{Recent work has introduced a ``not-quite-primordial" formation mechanism for SMBHs~\cite{Qin2025,Cyr2025} which can evade distortion constraints.}
For Gaussian fluctuations, it is well known that spectral distortion bounds from \textit{COBE/FIRAS} rule out power spectrum enhancements that could give rise to PBHs with initial mass $M_i \gtrsim 10^{4} \, M_{\odot}$ at formation (as was illustrated by the dashed orange curve in Fig.~\ref{fig:prk}) \cite{Chluba:2012we, Kohri:2014lza, Nakama:2017xvq, Karam:2022nym, Cyr2023}, inhibiting the production of seeds which could potentially grow to supermassive sizes by redshift $z \simeq 7$. This begs the question of how much non-Gaussianity one needs in order to evade $\mu$-distortion bounds and produce a black hole abundance that can be reconciled with the LRD mass function inferred by JWST. A self-consistent derivation of the correction to the $\mu$-distortion amplitude in the presence of local type non-Gaussianity has recently been presented \cite{Sharma:2024img}, which makes it possible to assess (in a Bayesian way) the allowed regions of parameter space for a variety of different power spectrum templates. In this section, we summarize this correction, following closely the formalism presented in Refs.~\cite{Sharma:2024img} and \cite{Wang:2025hbw}.

We note also a recent work which claimed that sufficiently non-Gaussian tails to the PDF could alleviate $\mu$-distortion constraints~\cite{Dave2026}. While certainly intriguing, the authors of that work did not consider the loop correction to the $\mu$-distortion\footnote{The authors also used an outdated version of the $\mu$ limit from \textit{COBE/FIRAS}, roughly a factor of 2 looser than what is considered here.} (discussed in Sec.~\ref{sec:level3.2}) that arises in the presence of non-Gaussian extensions \cite{Sharma:2024img}, nor did they compute their abundances using the compaction function approach (described in Sec.~\ref{sec:level2.1}), instead opting for a more simplified treatment (i.e. Eq.~\eqref{simplebeta}). Another article also appeared around the same time which did not suffer from these shortcomings, considering specifically the curvaton scenario~\cite{Allegrini:2026jqt}.

\subsection{\label{sec:level3.1}Gaussian perturbations}

Over the past decade, a simple analytic prescription has emerged which can reliably estimate the amplitude of a $\mu$-type distortion given a source of non-thermal energy and/or entropy injection. Generically, this can be written as \cite{Chluba2013a,Chluba2013b, Chluba2015b, Chluba2016, Cyr2023}
%
\begin{align} \label{eq:mu-greens}
    \mu \simeq 1.4 \int_0^{\infty} \id z \, \left(\frac{\id(Q/\rho_{\gamma})}{\id z} - \frac{4}{3} \frac{\id (N/n_{\gamma})}{\id z} \right) \mathcal{J}_{\mu}(z).
\end{align}
%
Here $Q/\rho_{\gamma}$ and $N/n_{\gamma}$ are the non-thermal energy and entropy (number density) injection rates relative to the CMB blackbody spectrum, and $\mathcal{J}_{\mu}(z)$ is the $\mu$ distortion visibility function, of which a number of different parameterized forms have been compared in previous work \cite{Cyr2023}. For the Silk damping distortion, non-thermal photons are neither created nor destroyed, as the diffusion proceeds solely through rapid Thomson interactions, thus $N = 0$ in our case. A precise computation of the heating rate \cite{Chluba2012a,Sharma:2024img} shows that
%
\begin{align}
    \frac{\id (Q/\rho_{\gamma})}{\id z} \simeq \frac{4 a \dot\tau}{H} \int_0^{\infty} \frac{\id k}{k} \mathcal{P}_{\mathcal{R}}(k) &\bigg[ \frac{(3 \Theta_1 - v)^2}{3} + \frac{9}{2} \Theta^2_2 \nonumber \\
    &- \frac{\Theta_2}{2}(\Theta_0^{\rm P} + \Theta_2^{\rm P})\bigg],
\end{align}
%
where $H$ is the Hubble rate, $a$ is the scale factor normalized to unity today, $\mathcal{P}_{\mathcal{R}} (k)$ is the dimensionless primordial power spectrum, $\dot{\tau} = \sigma_{\rm T} N_{\rm e}$ is the Thomson scattering rate (with $N_{\rm e}$ being the free electron density), and $v$ is the baryon velocity. The $\Theta_N$ refer to the $N$th moment of the photon temperature fluctuation (e.g. $N=0,1,2$ are the monopole, dipole, and quadrupole respectively), with a superscript ${\rm P}$ referring to polarization fluctuations. When one restricts to the tight coupling limit \cite{Hu1995} (which certainly applies during the $\mu$ era), this rate simplifies to
%
\begin{align} \label{eq:SD-tight}
    \frac{\id (Q/\rho_{\gamma})}{\id z} \simeq \frac{32}{45 \dot\tau}\frac{A^2}{H a} \int_0^{\infty} \id k \,\, k \,\mathcal{P}_{\mathcal{R}}(k) \, {\rm e}^{-(k/k_{\rm D})^2}.
\end{align}
%
To account for isocurvature fluctuations \cite{Chluba2013}, the amplitude of the WKB mode ($A$) receives $k$-dependent corrections, whereas for adiabatic modes $A \simeq 0.9$. The damping scale $k_{\rm D}$ can be approximated \cite{Hu1995, Kosowsky1995,Dodelson2020} by
%
\begin{align}
    k_{\rm D} \simeq 2\pi \left(\int \id z \, \frac{c_{\rm s}^2}{2 \dot{\tau} H} \left[ \frac{R^2}{1+R} + \frac{16}{15} \right] \right)^{-1/2},
\end{align}
%
with $R = 3\rho_{\rm b}/4\rho_{\gamma}$ being the baryon loading, and $c_{\rm s}$ the sound speed. This quantity is redshift dependent and describes the time at which a given $k$-mode deposits most of its energy into the background, producing a distortion. A back-of-the-envelope estimation of this yields $z_{\rm dis} \simeq 4.5 \times 10^5 (k/10^3 \, {\rm Mpc}^{-1})^{2/3}$. Insertion of Eq.~\eqref{eq:SD-tight} into Eq.~\eqref{eq:mu-greens} provides a straightforward way to compute the expected $\mu$ distortion for any power spectrum (obeying Gaussian statistics), and inserting Planck-inferred values yields $\mu \simeq 2 \times 10^{-8}$ \cite{Chluba2012a, Cyr2023, Cyr_SD_whitepaper}. In the limit of small distortions ($\mu \ll 1$), the background cosmology (which sets quantities like the damping scale) is not expected to change dramatically, thus it is common to write the distortion as
%
\begin{align} \label{eq:mu_estimate}
    \mu \simeq \int \frac{\id k}{k} \, \mathcal{P}_{\mathcal{R}}(k) \, W_{\mu} (k),
\end{align}
%
where the integration over $z$ of the background quantities is packaged into the so-called $\mu$-distortion window function $W_\mu (k)$ for the damping of acoustic modes. Similar to the distortion visibility function $\mathcal{J}_{\mu}$, various approximations for $W_{\mu}(k)$ have recently been computed and compared~\cite{Cyr2023, Sharma:2024img}. Using current cosmological parameter values, Sharma \textit{et al.}~\cite{Sharma:2024img} used CLASS \cite{Lesgourgues2011,Blas2011} to compute the distortion (for a peaked power spectrum), and performed a PCA decomposition to determine
%
\begin{align}
    W_{\mu}(k) = {\rm exp}\left[\sum_{n=0}^6 \, w_n \left(\ln \left[ \frac{k}{1 \, {\rm Mpc}^{-1}}\right] \right)^n \right],
\end{align}
%
where the coefficients $w_{n}$ are given in Table I of Ref.~\cite{Sharma:2024img}. In this work, we use this expression when estimating the $\mu$ distortion.

\subsection{\label{sec:level3.2} Loop correction to {\boldmath$\mu$} from non-Gaussianity}

Whereas Eq.~\eqref{eq:mu_estimate} is sufficient for (adiabatic) Gaussian perturbations, it requires correcting when primordial non-Gaussianity is present. Intuitively, the inclusion of non-Gaussianity will generally yield an increase in the expected $\mu$ distortion relative to its Gaussian counterpart when the tail of the probability distribution rises. 

Non-Gaussianity implies that the background perturbations are not just described by the power spectrum, but also by higher-point correlation functions such as the bispectrum. For the case of local type non-Gaussianity, the (position space) comoving curvature perturbation can be expressed as in Eq.~\eqref{eq:Rform}. Recent work \cite{Sharma:2024img} has shown that for $|\tilde{f}_{\rm NL}| > 1$, the correction to $\mu$ coming from the bispectrum (and higher-point correlations) are subdominant to the piece coming from the modification of the power spectrum. \textit{COBE/FIRAS} provides a very liberal constraint of $\tilde{f}_{\rm NL} \lesssim 2 \times 10^6$ over these small scales \cite{Bianchini:2022dqh}. 

We are primarily interested in large deviations away from Gaussianity, and so for simplicity we consider only the correction to $\mu$ coming from the modified power spectrum. In this case, it is known \cite{Byrnes2007} that the (dimensionful) power spectrum acquires a 1-loop contribution of the form
%
\begin{align}
    P_{\mathcal{R}}(k) &= P_{\mathcal{R}_{\rm G}}(k) + \tilde{f}^2_{\rm NL}P_{\mathcal{R}_{\rm G}^2}(k) \nonumber \\
    &= P_{\mathcal{R}_{\rm G}}(k) + 2 \tilde{f}_{\rm NL}^2 \int \frac{\id^3 q}{(2\pi)^3} P_{\mathcal{R}_{\rm G}}(q) \, P_{\mathcal{R}_{\rm G}}(|\vec{k} - \vec{q}|).
    \label{eq:prk-ng-loop}
\end{align}
%
The dimensionless and dimensionful power spectra are related through $\mathcal{P_{\mathcal{R}_{\rm G}}} = (k^3/2\pi^2) P_{\mathcal{R}_{\rm G}}$. Inserting this into Eq.~\eqref{eq:mu_estimate} yields the schematic form
%
\begin{align}\label{eq:mu-total}
    \mu &= \mu_{\rm tree} + \mu_{\rm loop} \nonumber \\
    &= A \, \mathcal{W}_{\mu}^{\rm (G)} + 2 \tilde{f}_{\rm NL}^2 A^2 \mathcal{W}_{\mu}^{\rm (NG)}.
\end{align}
%
In this case, we have assumed that the amplitude of the power spectrum can be factorized, e.g. $P_{\mathcal{R}_{\rm G}}(k,A, ...) = A \,  \tilde{P}_{\mathcal{R}_{\rm G}}(k,...)$ where $(...)$ refers to the additional parameters such as the peak scale $k_p$ in the case of a delta function. The problem then reduces to needing to calculate the integrated Gaussian and non-Gaussian window functions, 
%
\begin{align}
    \mathcal{W}_{\mu}^{\rm (G)}(...)  &=\int \id k \frac{k^2}{2\pi^2} W_{\mu}(k) \tilde{P}_{\mathcal{R}_{\rm G}}(k,...), \\
    \mathcal{W}_{\mu}^{\rm (NG)}(...)  &=\int \id k \frac{k^2}{2\pi^2} W_{\mu}(k) \nonumber \\
    &\times \int \frac{\id^3 q}{(2\pi)^3} \tilde{P}_{\mathcal{R}_{\rm G}}(q,...) \tilde{P}_{\mathcal{R}_{\rm G}}(|\vec{k}-\vec{q}|,...).
\end{align}
%
While the computation of $\mathcal{W}_{\mu}^{\rm (G)}$ is generally straightforward (and sometimes analytic), determining $\mathcal{W}_{\mu}^{\rm (NG)}$ can be more numerically costly. Nevertheless, we have developed efficient numerical routines to implement it into the MCMC analysis, allowing us to self-consistently compute the one-loop contribution to the $\mu$ distortion in our results.

In principle, one could consider the presence of higher-order terms in the non-Gaussian expansion, for example coming from $g_{\rm NL}$. The correction to $\mu$ from this would be captured by a corresponding two-loop diagram. This has been done recently where it was shown that (for peaked power spectra) even for extreme values of $g_{\rm NL}$, the change in the $\mu$ distortion constraint is marginal~\cite{Pritchard2025}. We leave the inclusion of these effects on more broad power spectra to future work.

\section{\label{sec:methods} Data Sets and Methodology}

We employ two separate observational probes, namely, CMB $\mu$-type spectral distortions and LRD number density, to constrain the primordial power spectrum on small scales. 

\begin{itemize}
    \item {\bf CMB Spectral Distortions:} We use the FIRAS measurement of the CMB $\mu$-distortion (Ref.~\cite{Fixsen1996}) and
    adopt the FIRAS $95\%$ confidence interval (CI) upper limit of $|\mu|<4.7\times 10^{-5}$ \cite{Bianchini:2022dqh}.
    The likelihood for the $\mu$-distortion constraint is approximated by a Gaussian with mean $\mu_{\rm obs}=0$ and standard deviation $\sigma_{\mu}=2.4\times10^{-5}$. 
    As discussed in Sec.~\ref{sec:level3}, we include both the tree-level and loop contributions to $\mu$. 
    
    \item \textbf{JWST LRDs:} Following the analysis of Ref.~\cite{2024ApJ...968....4P}, we take the observed number density for LRDs to be $\log_{10}(n_{\rm LRD} \, {\rm Mpc}^{-3}/n_{\rm ref})=-4.0\pm 0.1$, where $n_{\rm ref}=1 \, {\rm Mpc^{-3}}$ is a reference comoving number density. We then construct the likelihood as lognormal in $n_\mathrm{LRD}$, where $n_\mathrm{LRD}$ is determined by integrating over black holes at $z_\mathrm{LRD} = 7$ with masses greater than $10^6 \,M_\odot$.
\end{itemize}

\noindent In all analyses the PBH abundance is calculated using the compaction function formalism described in Sec.~\ref{sec:level2} and the $\mu$-distortion is calculated from Eq.~\eqref{eq:mu-total}.

We perform Bayesian parameter estimation using the affine-invariant ensemble sampler \texttt{emcee} \cite{Foreman_Mackey_2013}. 
Each analysis utilizes an ensemble of 200 walkers. For MCMCs constrained only by CMB $\mu$-distortions, all analyses converged within 5,000 independent steps. For those analyses constrained by both $\mu$-distortion constraints and LRD number density, we monitor convergence using the autocorrelation time $\tau$.
To limit computational expense, we do not enforce a convergence criterion on $\tau$ directly, but rather monitor the stability of posterior means and bounds.  Analyses for $f_{\rm duty}=1$ with perturbative priors completed a minimum of 8,000 steps per walker, while analyses with $f_{\rm duty}= 0.5, 0.3, 0.1$ completed 20,000  steps. MCMCs with flat priors completed a minimum of 2,000 steps.

We sample the $\prk$ template parameters: the peak wavenumber $k_p$, the amplitude at the spectral peak $A$ (and the dimensionless peak width, $\Delta$, in the lognormal template), and the local non-Gaussianity $\tilde{f}_{\rm NL}$. We adopt broad uniform priors over these parameters, subject to an upper limit $A<10^{-1.5}$ to parametrize bounds on PBHs reviewed in Ref.~\cite{Carr:2020gox}, and a perturbativity prior that excludes regions of the parameter space in which the local non-Gaussian expansion in the Gaussian curvature, $F({\cal R}_{G})=\mathcal{R}_G+\tilde{f}_{\rm NL}\mathcal{R}_G^2$ fails to converge \cite{Iovino2024}. 
The perturbative constraint takes the form 
%
\begin{equation}
    |\tilde{f}_{\rm NL}|\cdot\sigma_{r}(A,\Delta_{\rm shape}) < 1/4 ,\label{eq:pertprior_general}
\end{equation}
%
and depends implicitly on the shape of the power spectrum through a spectrum-dependent shape parameter, $\Delta_{\rm shape}$, which enters through the variance $\sigma_r^2$. 
The shape parameter $\Delta_{\rm shape}$ is $\Delta$ (the peak width) and $n_*$ (the spectral tilt) for the lognormal and cut-power-law templates respectively, as defined in Eqs.~(\ref{lognormalPR}) and (\ref{sharkPR}). 
The variance is an integral of the (dimensionful) curvature power spectrum $P_{\cal R}(k)$ against the window and transfer functions $W_s,\,W,\,T$ (given in Eq.~\eqref{eq:transfer-functions}) evaluated at the value of the characteristic smoothing scale $r_m$ that maximizes the total compaction. 
The dimensionful power spectrum factorizes as $P_{\cal R}(k)= A {\tilde P}_{\cal R}(k)$ into an amplitude and dimensionless power spectrum, which contains the dependence on the shape parameter $\Delta_{\rm shape}$. This factorization is then inherited by the variance, so that we may write $\sigma_{r}^2$ in terms of a dimensionless correlator per unit amplitude $\hat{\sigma}_{r}^2(\Delta_{\rm shape})=\sigma_{r}^2(\Delta_{\rm shape})/A$. The quantities $\sigma_{r}^2(\Delta_{\rm shape})/A$ are computed once from an interpolation built from a fixed set of values, following Ref.~\cite{Sharma:2024img}, and tabulated. 

In the cut-power-law template, we fix the shape parameter, $n_*=4$, so that $\sigma_{r}^2/A=0.593$ takes only a single value. Rewriting the constraint \eqref{eq:pertprior_general} in terms of $\hat{\sigma}_{r}=\sqrt{\sigma_r^2/A}$ gives 
%
\begin{align}
          |\tilde{f}_{\rm NL}|\sqrt{A}< 0.325 & \text{\,\, (CPL)}\label{priorCPL},\\
    |\tilde{f}_{\rm NL}|\sqrt{A}<\frac{1}{4\hat{\sigma}_r(\Delta)} & \text{\,\, (LN).}\label{priorLN}
\end{align}
%
Physically, the intrinsically joint prior over the peak amplitude and local non-Gaussianity enforces that the non-Gaussian correction to the two-point correlation function $2\tilde{f}_{\rm NL}^2 \sigma_{r}^4$ remain strongly subdominant to the Gaussian part $\sigma_r^2$: the non-Gaussian correction can be at most about $12\%$ of the Gaussian part. 

A comparison of constraints with and without the perturbative non-Gaussianity prior is given in Appendix~\ref{app:supp_plots} in Fig.~\ref{fig:priors-comp}.

\section{\label{sec:results} Constraints from Data}

The main results are shown in Figs.~\ref{fig:main-corner-ln} to \ref{fig:fduty-full}, with supplementary figures given in Appendix~\ref{app:supp_plots}. Marginalized parameter constraints are tabulated in Tables~\ref{tab:ln-results} to \ref{tab:cpl-results-fduty} for the lognormal and cut-power-law templates.

\begin{figure*}
    \centering
    \includegraphics[width=\linewidth]{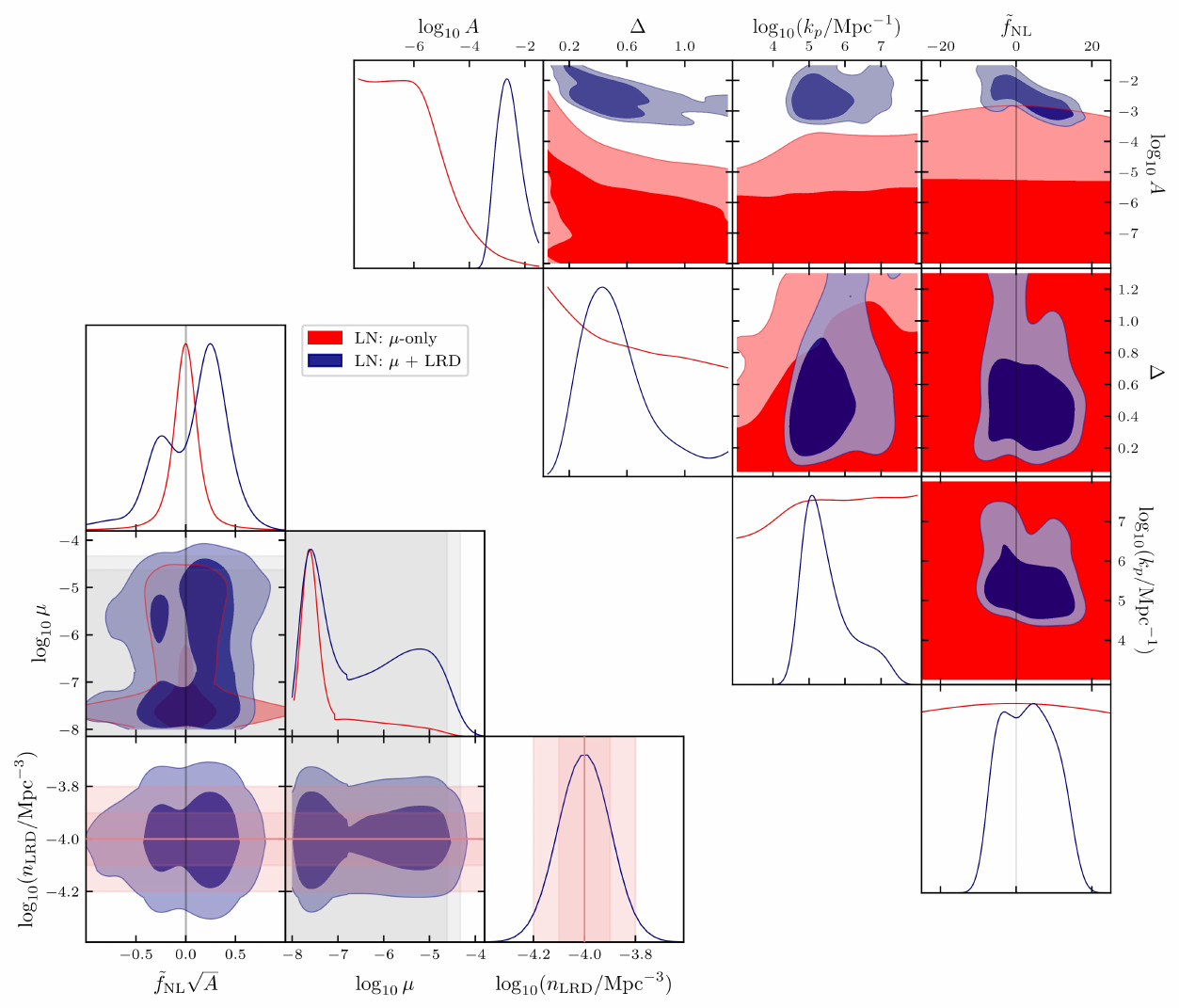}
    \caption{Constraints on the lognormal (LN) power spectrum profile from $\mu$-only (red) and $\mu$+LRD (navy) likelihoods. We adopt a perturbative non-Gaussianity prior and assume continuous Eddington accretion (i.e. $f_{\rm duty} = 1$). We impose an upper limit $A<10^{-1.5}$ to parametrize bounds on PBHs reviewed in Ref.~\cite{Carr:2020gox}.
    The $1\sigma$ ($2\sigma$) constraints on $\mu$ and $n_{\rm LRD}$ are indicated by the light (dark) gray and light (dark) red shaded bands, respectively, with the red line representing the mean of the observed LRD abundance, $\log_{10}(n_{\rm LRD} / {\rm Mpc}^{-1})=-4.0$. The gray line highlights the Gaussian case ($\tilde{f}_{\rm NL}=0$).
    The bottom left panels show how the main observables ($\mu$ and $n_{\rm LRD}$) depend on the reduced model parameter $\tilde{f}_{\rm NL} \sqrt{A}$, and that they correctly fit observational constraints.
    The top right panels show the posteriors for the model parameters $A$, $\Delta$, $k_p$, and $\tilde{f}_{\rm NL}$. 
    }
    \label{fig:main-corner-ln}
\end{figure*}

\subsection{Constraints with Continuous Eddington Accretion ({\boldmath$f_{\rm duty}=1$})}\label{sec:results:fduty=1}

\subsubsection{Lognormal Power Spectrum}\label{sec:results:fduty=1:LN}

We begin by considering the lognormal power spectrum, Eq.~\ref{lognormalPR}, with perturbative priors and accretion modeled by continuous Eddington accretion ($f_{\rm duty}=1$). The posterior distributions of the MCMCs are shown in Fig.~\ref{fig:main-corner-ln} and the derived parameter constraints are given in Tab.~\ref{tab:ln-results}. We analyze the model given the \textit{COBE/FIRAS} $\mu$-distortion upper bound, both with (``$\mu$+LRD'', navy) and without (``$\mu$-only'', red) the requirement that the model also explains the LRD population observed by JWST. In Fig.~\ref{fig:main-corner-ln}, the top (inverted) corner plot shows the posteriors of the model parameters $A$, $\Delta$, $k_p$, and $\tilde{f}_{\rm NL}$, while the bottom corner plot shows those of the combined model parameter $\tilde{f}_{\rm NL}\sqrt{A}$ together with the observational results, $\mu$ and $n_{\rm LRD}$.

\begin{table}[h!]
Constraints on a lognormal power spectrum with perturbative non-Gaussianity. 
    \centering
    \begin{tabular} {|l|c|c|}
        \hline\hline 
        \rule{0pt}{1em} Parameter & \textit{COBE/FIRAS} $\mu$ & + JWST LRD 
        \\ \hline\hline
        \rule{0pt}{1em}
        {\boldmath$\log_{10} A$} 
        & $-6.2(-4.0)^{+0.5}_{-1.8}$ 
        & $-2.57(-2.62)^{+0.36}_{-0.53}$
        \\
        \rule{0pt}{1em}
        {\boldmath$\Delta$} 
        & $0.62(0.61)^{+0.22}_{-0.57}$ 
        & $0.53(0.61)^{+0.11}_{-0.31}$
        \\
        \rule{0pt}{1em}
        {\boldmath$\log_{10}k_p$} 
        & $5.6(7.4)^{+2.4}_{-2.6}$ 
        & $5.50(6.26)^{+0.25}_{-0.85}$
        \\
        \rule{0pt}{1em}
        {\boldmath$\tilde{f}_{\rm NL}$} 
        & $-1(0)\pm{120}$ 
        & $3.1(-7.9)^{+5.7}_{-9.5}$
        \\
        \hline
        \rule{0pt}{1em}
        $\tilde{f}_{\rm NL} \sqrt{A}$ 
        & $-0.005(0)\pm{0.11}$ 
        & $0.05(-0.39)^{+0.41}_{-0.40}$
        \\
        \rule{0pt}{1em}
        $\log_{10}\mu$
        & $-7.22(-7.64)^{+0.002}_{-0.42}$ 
        & $-6.4(-7.6)^{+1.8}_{-1.4}$
        \\
        \rule{0pt}{1em}
        $\log_{10}\mu_{\rm loop}$
        & $-10.3(-21.0)^{+0.7}_{-1.9}$ 
        & $-7.9(-9.6)^{+2.6}_{-0.9}$
        \\
        \rule{0pt}{1em}
        $\log_{10}n_{\rm LRD}$
        & ---
        & $-4.00(-4.00)\pm 0.10$
        \\
        \hline
    \end{tabular}
    \caption{Marginalized parameter constraints and best-fit values on a lognormal power spectrum from \textit{COBE/FIRAS} $\mu$-distortion constraints alone and in combination with JWST LRDs. Best-fit values are approximated by the best sample in the chain. For each parameter, the constraints are given as ``mean (best-fit) $\pm$ std.'' The parameters $k_p$ and $n_{\rm LRD}$ are given in units of (comoving) ${\rm Mpc}^{-1}$ and ${\rm Mpc}^{-3}$, respectively.}
    \label{tab:ln-results}
\end{table}

\begin{figure*}
    \centering
    \includegraphics[width=\linewidth]{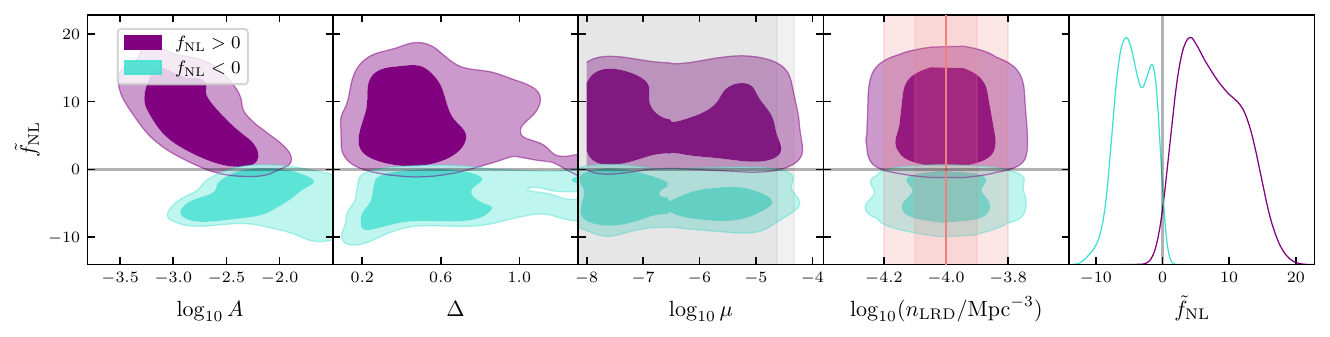}
    \caption{Bimodal structure of $\tilde{f}_{\rm NL}$: the posterior distribution of $\tilde{f}_{\rm NL}$ splits neatly into two lobes with distinct peaks for positive and negative $\tilde{f}_{\rm NL}$. Shown here is the $\tilde{f}_{\rm NL}$ posterior of the lognormal power spectrum profile with perturbative priors and $\mu$+LRD likelihood. 
    The $1\sigma$ ($2\sigma$) constraints on $\mu$ and $n_{\rm LRD}$ are indicated by the light (dark) gray and light (dark) red shaded bands, respectively, with the red line representing the mean of the observed LRD abundance, $\log_{10}(n_{\rm LRD} / {\rm Mpc}^{-1})=-4.0$.
    We note that the support for crossing the boundary at $\tilde{f}_{\rm NL}=0$ (indicated by the gray line) is an artifact of smoothing. 
    } 
    \label{fig:fnl-branches}
\end{figure*}

The constraint from the $\mu$-distortions alone is shown in the red contours in Fig.~\ref{fig:main-corner-ln}, and the middle column of Tab.~\ref{tab:ln-results}. It may be appreciated from the $\tilde{f}_{\rm NL}\sqrt{A}-\mu$ panel (center-left panel in the bottom corner plot) that the model is largely able to satisfy the $\mu$-distortion constraints: the vast majority of the red contour sits within the gray shading representing the \textit{COBE/FIRAS} constraint. Notably, the model exhibits strong support for very small $\mu$ values, which reduce to the $\Lambda$CDM floor.

The top right panels of Fig.~\ref{fig:main-corner-ln} indicate that the model satisfies the $\mu$-distortion constraints for a wide range of the model parameters: the posteriors nearly saturate the prior parameter range. In the absence of LRD constraints, the MCMC posteriors indicate a very mild preference for $\tilde{f}_{\rm NL}=0$, i.e. Gaussian primordial perturbations. 

These constraints change dramatically when LRD constraints are included, as shown by the navy contours of Fig.~\ref{fig:main-corner-ln} and the right-hand column of Tab.~\ref{tab:ln-results}. Compared to the $\mu$-only constraints (in red), the top corner plot highlights a significant restriction of the posterior range. Notably, in order to produce the LRD population, a strong enhancement of the curvature power spectrum becomes necessary: the peak amplitude $A$ is pushed to large values (see top column). To maintain consistency with the $\mu$-distortion constraints, the location of the peak is also shifted towards larger $k_p$ values.
Furthermore, one may now appreciate a preference for non-zero $\tilde{f}_{\rm NL}$ (see marginalized $\tilde{f}_{\rm NL}$ distribution in the bottom panel).
We note however that the Gaussian case ($\tilde{f}_{\rm NL}=0$) is not excluded by LRDs; this reflects the assumption of continuous Eddington accretion, which we will return to in Sec.~\ref{sec:results:fduty<1}. 

The $\mu$+LRD constraints can be understood by referring back to Fig.~\ref{fig:prk}. In order to produce PBHs, a large enhancement of $\prk (k)$ is required. For Gaussian perturbations and inefficient accretion ($f_{\rm duty} \sim 0$), such an enhancement is prohibited by the \textit{COBE/FIRAS} bound for $k \lesssim 10^5 \, {\rm Mpc}^{-1}$. 
(At larger scales of $k \lesssim {\rm Mpc}^{-1}$ even tighter constraints can be set by large scale structure and CMB anisotropy constraints.) Incorporation of non-Gaussianity facilities the production of PBHs for reduced power-spectrum amplitudes; together with efficient accretion (with $f_{\rm duty} \sim 1$), these features can produce a population of SMBH seeds that yield the observed LRD population by $z \sim 7$.

Consistent with this, the $\tilde{f}_{\rm NL}\sqrt{A}-n_{\rm LRD}$ posterior in Fig.~\ref{fig:main-corner-ln} indicates that the production of LRD-compatible PBH seeds favors non-zero $\tilde{f}_{\rm NL}$ values. 
This is accompanied by an increased preference for values of $\mu$ near the $COBE/FIRAS$ upper bound, as can be appreciated from the $\tilde{f}_{\rm NL}\sqrt{A}-\mu$ panel Fig.~\ref{fig:main-corner-ln}.
Overall, these constraints demonstrate that
it is possible to match the LRD number density observed by JWST while remaining below the \textit{COBE/FIRAS} $\mu$-constraints. This is captured the $\mu-n_{\rm LRD}$ panel, which shows near-perfect mutual agreement between the model posteriors and the two bounds.

\begin{table}[h!]
Constraints on the positive and negative $\tilde{f}_{\rm NL}$ modes. 
    \centering
    \begin{tabular} {|l|c|c|}
        \hline\hline 
        \rule{0pt}{1em} Parameter & Positive $\tilde{f}_{\rm NL}$  & Negative $\tilde{f}_{\rm NL}$ 
        \\ \hline\hline
        \rule{0pt}{1em}
        {\boldmath$\log_{10} A$} 
        & $-2.76(-3.22)^{+0.30}_{-0.39}$ 
        & $-2.24(-2.62)^{+0.40}_{-0.39}$
        \\
        \rule{0pt}{1em}
        {\boldmath$\Delta$} 
        & $0.50(0.98)^{+0.09}_{-0.27}$ 
        & $0.58(0.61)^{+0.12}_{-0.37}$
        \\
        \rule{0pt}{1em}
        {\boldmath$\log_{10}k_p$} 
        & $5.31(6.17)^{+0.23}_{-0.70}$ 
        & $5.81(6.26)^{+0.40}_{-0.98}$
        \\
        \rule{0pt}{1em}
        {\boldmath$\tilde{f}_{\rm NL}$} 
        & $7.5(6.5)^{+3.5}_{-6.0}$ 
        & $-4.3(-7.9)^{+3.6}_{-2.9}$
        \\
        \hline
        \rule{0pt}{1em}
        $\tilde{f}_{\rm NL} \sqrt{A}$ 
        & $0.29(0.16)^{+0.11}_{-0.18}$ 
        & $-0.35(-0.39)^{+0.30}_{-0.05}$
        \\
        \rule{0pt}{1em}
        $\log_{10}\mu$
        & $-6.3(-7.6)^{+1.8}_{-1.5}$ 
        & $-6.6(-7.6)^{+1.8}_{-1.1}$
        \\
        \rule{0pt}{1em}
        $\log_{10}\mu_{\rm loop}$
        & $-7.6(-9.9)^{+2.2}_{-0.8}$ 
        & $-8.4(-9.6)^{+3.1}_{-1.8}$
        \\
        \rule{0pt}{1em}
        $\log_{10}n_{\rm LRD}$
        & $-4.00(-4.00)\pm 0.10$
        & $-4.00(-4.00)\pm 0.10$
        \\
        \hline
    \end{tabular}
    \caption{
    Marginalized parameter constraints and best-fit values on a lognormal power spectrum from \textit{COBE/FIRAS} $\mu$-distortion constraints in combination with JWST LRDs, for the positive versus negative modes of $\tilde{f}_{\rm NL}$. 
    Best-fit values are approximated by the best sample in the chain. For each parameter, the constraints are given as ``mean (best-fit) $\pm$ std.'' The parameters $k_p$ and $n_{\rm LRD}$ are given in units of (comoving) ${\rm Mpc}^{-1}$ and ${\rm Mpc}^{-3}$, respectively.
    }
    \label{tab:fnl-results}
\end{table}

\begin{figure}
    \centering
    \includegraphics[width=.55\linewidth]{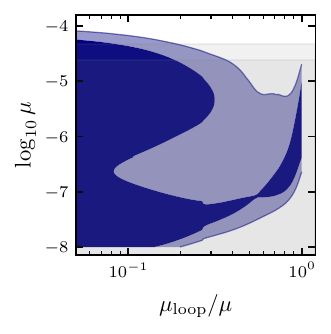}
    \caption{Loop correction to $\mu$: Joint posterior distribution for the total $\mu$ distortion and relative contribution from the loop correction, $\mu_{\rm loop}/\mu$, for the lognormal power spectrum profile with perturbative priors and $\mu$+LRD likelihood. 
    The $1\sigma$ ($2\sigma$) constraints on $\mu$ are indicated by the light (dark) gray shaded bands.
    For small $\mu$, far from the \textit{COBE/FIRAS} bound, $\mu_{loop}$ can be an ${\cal O}(1)$ contribution to the total $\mu$-distortion, while for large $\mu$, close the bound, the ratio $\mu_{\rm loop}/\mu$ is at most ${\cal O}(0.1)$. See also Fig.~\ref{fig:mu} in Appendix~\ref{app:supp_plots}.
    }
    \label{fig:mu-square}
\end{figure}

Interestingly, it may be noticed from the navy contours of Fig.~\ref{fig:main-corner-ln} that, in the $\mu$+LRD case, a bimodal distribution for $\tilde{f}_{\rm NL}$ emerges. This is consistent with Fig.~\ref{fig:fNLnLRD} of Sec.~\ref{sec:LRD}, which demonstrated that both negative and positive $\tilde{f}_{\rm NL}$ can seed LRDs in adequate abundance. 
To better illustrate this bimodality, we divide the sample into positive-$\tilde{f}_{\rm NL}$ and negative-$\tilde{f}_{\rm NL}$ modes, as shown in Fig.~\ref{fig:fnl-branches}. The corresponding constraints are summarized in Tab.~\ref{tab:fnl-results}. Both modes allow for the production of PBH seeds consistent with the JWST LRD population and the \textit{COBE/FIRAS} $\mu$-constraints. In both cases, smaller values of $\vert \tilde{f}_{\rm NL}\vert$ require larger local enhancements of the curvature power spectrum, as quantified by $A$, to produce sufficient PBH seeds. 
Note that, for both modes, $|\tilde{f}_{\rm NL}|$ always remains of ${\cal O}(1)$.

It is also informative to analyze the impact of non-Gaussianities on the $\mu$-distortion signal, which enters through the $\mu_{\rm loop}$ term of Eq.~\eqref{eq:mu-total}.  Fig.~\ref{fig:mu-square} shows the 2d posterior for the total $\mu$-distortion and the relative loop contribution $\mu_{\rm loop}/\mu$, for the lognormal power spectrum with perturbative priors, in the $\mu$+LRD scenario. (Additional panels are shown in Fig.~\ref{fig:mu}). From Fig.~\ref{fig:mu-square} it can be appreciated that for $\mu$ far from the \textit{COBE/FIRAS} bound, $\mu_{\rm loop}$ can be an ${\cal O}(1)$ contribution to the total $\mu$-distortion, whereas for $\mu$ close to the bound, the ratio $\mu_{\rm loop}/\mu$ is at most ${\cal O}(0.1)$.

\begin{figure}
    \centering
    \includegraphics[width=.55\linewidth]{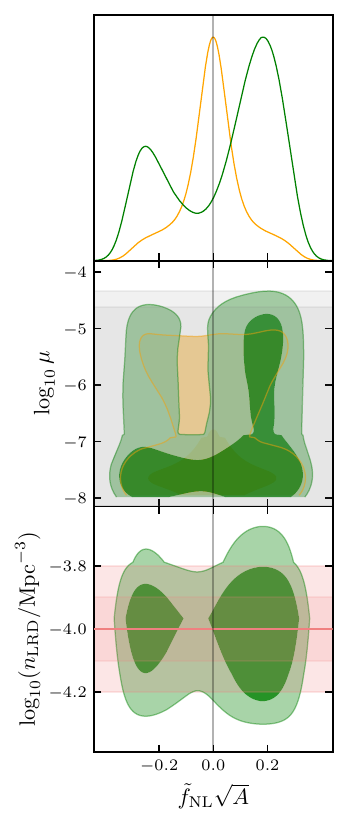}
    \caption{Cut-power-law (CPL) model: Constraints on the cut-power-law power spectrum from $\mu$ (yellow) and $\mu$+LRD (green) likelihoods. We adopt a perturbative non-Gaussianity prior and assume continuous Eddington accretion ($f_{\rm duty}=1$). 
    The $1\sigma$ ($2\sigma$) constraints on $\mu$ and $n_{\rm LRD}$ are indicated by the light (dark) gray and light (dark) red shaded bands, respectively, with the red line representing the mean of the observed LRD abundance, $\log_{10}(n_{\rm LRD} / {\rm Mpc}^{-1})=-4.0$.
    We note that posterior support at $\tilde{f}_{\rm NL}=0$ (indicated by the gray line) reflects that continuous Eddington accretion ($f_{\rm duty}=1$) can satisfy both $\mu$ and LRD targets even for $\tilde{f}_{\rm NL}=0$. Full parameter constraints are described in Tab.~\ref{tab:cpl-results} and also shown in Appendix~\ref{app:supp_plots}, see Fig.~\ref{fig:main-corner-cpl}.}
    \label{fig:cpl-column}
\end{figure}

\begin{figure*}
    \centering
    \includegraphics[width=\linewidth]{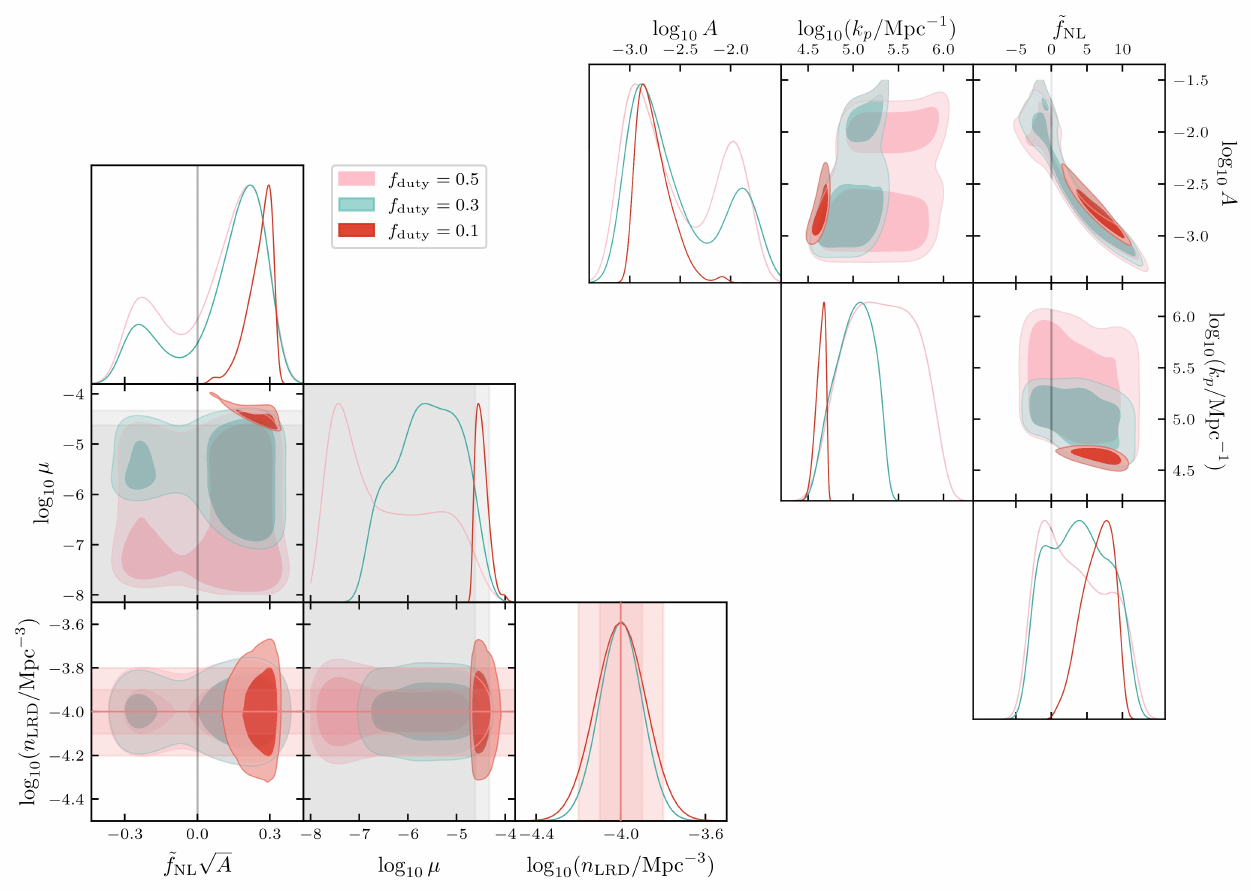}
    \caption{Varying duty cycles: Constraints from $\mu$ and LRD for varying accretion duty cycle parameter $f_{\rm duty}$. We consider a cut-power-law power spectrum  with perturbative priors and $\mu$+LRD likelihoods, for varying values of $f_{\rm duty}$. 
    The $1\sigma$ ($2\sigma$) constraints on $\mu$ and $n_{\rm LRD}$ are indicated by the light (dark) gray and light (dark) red shaded bands, respectively, with the red line representing the mean of the observed LRD abundance, $\log_{10}(n_{\rm LRD} / {\rm Mpc}^{-1})=-4.0$. The gray line highlights the Gaussian case ($\tilde{f}_{\rm NL}=0$).
    The bottom corner plot shows how the main observables ($\mu$ and $n_{\rm LRD}$) depend on the reduced model parameter $\tilde{f}_{\rm NL} \sqrt{A}$, and that they correctly fit observational constraints. The top corner plot shows the posteriors for the model parameters $A$, $k_p$, and $\tilde{f}_{\rm NL}$.}
    \label{fig:fduty-full}
\end{figure*}

\subsubsection{Cut-Power-Law Power Spectrum}\label{sec:results:fduty=1:CPL}

We repeat the full analysis of \S\ref{sec:results:fduty=1:LN} for the cut-power-law power spectrum template of Eq.~\eqref{sharkPR}. We show the posteriors for the model in the fit to the $\mu$-only and the $\mu$+LRD likelihoods in Fig.~\ref{fig:cpl-column}, with more complete results shown in Appendix~\ref{app:supp_plots} in Fig.~\ref{fig:main-corner-cpl}, and give the corresponding parameter constraints in Tab.~\ref{tab:cpl-results}. Fig.~\ref{fig:cpl-column} shows a subset of the constraints while a full constraints plot is shown in Fig.~\ref{fig:main-corner-cpl}. The constraints mirror the trends shown for the lognormal case, with the LRD constraints picking out two distinct modes of $\tilde{f}_{\rm NL}$. The model can easily accommodate $\mu$-distortion constraints and produce adequate LRDs for $\tilde{f}_{\rm NL} = {\cal O}(1)$ and $A = {\cal O}(10^{-3})$.

\begin{table}[h!]
Constraints on a cut-power-law power spectrum with perturbative non-Gaussianity. 
    \centering
    \begin{tabular} {|l|c|c|}
        \hline\hline 
        \rule{0pt}{1em} Parameter & \textit{COBE/FIRAS} $\mu$ & + JWST LRD 
        \\ \hline\hline
        \rule{0pt}{1em}
        {\boldmath$\log_{10} A$} 
        & $-6.3(-6.5)^{+0.7}_{-1.5}$ 
        & $-2.59(-2.94)^{+0.59}_{-0.56}$
        \\
        \rule{0pt}{1em}
        {\boldmath$\log_{10}k_p$} 
        & $5.5(7.7){\pm 2.5}$ 
        & $5.8(6.8)^{+0.5}_{-1.1}$
        \\
        \rule{0pt}{1em}
        {\boldmath$\tilde{f}_{\rm NL}$} 
        & $-2(0)\pm{120}$ 
        & $2.7(5.6)^{+3.3}_{-6.5}$
        \\
        \hline
        \rule{0pt}{1em}
        $\tilde{f}_{\rm NL} \sqrt{A}$ 
        & $-0.001(0)\pm{0.095}$ 
        & $0.04(0.19)^{+0.25}_{-0.34}$
        \\
        \rule{0pt}{1em}
        $\log_{10}\mu$
        & $-7.27(-7.64)^{+0.06}_{-0.38}$ 
        & $-6.9(-7.6)^{+2.2}_{-0.8}$
        \\
        \rule{0pt}{1em}
        $\log_{10}\mu_{\rm loop}$
        & $-10.4(-19.2)^{+0.7}_{-1.7}$ 
        & $-9.4(-12.5)^{+3.3}_{-1.6}$
        \\
        \rule{0pt}{1em}
        $\log_{10}n_{\rm LRD}$
        & ---
        & $-4.00(-4.00)\pm 0.10$
        \\
        \hline
    \end{tabular}
    \caption{
    Marginalized parameter constraints and best-fit values on a cut-power-law power spectrum from \textit{COBE/FIRAS} $\mu$-distortion constraints alone and in combination with JWST LRDs. Best-fit values are approximated by the best sample in the chain. For each parameter, the constraints are given as ``mean (best-fit) $\pm$ std.'' The parameters $k_p$ and $n_{\rm LRD}$ are given in units of (comoving) ${\rm Mpc}^{-1}$ and ${\rm Mpc}^{-3}$, respectively.
    }
    \label{tab:cpl-results}
\end{table}

\subsubsection{Prior Dependence}

In Appendix~\ref{app:supp_plots}, we study the impact of the choice of priors on these constraints. We repeat the analysis of the lognormal model but with uniform priors on the non-Gaussianity instead of a perturbative prior. Results are shown in Fig.~\ref{fig:priors-comp}. The qualitative results are essentially unchanged, while the uniform priors show stronger preference for $\mu$ near the \textit{COBE/FIRAS} upper bound.

\subsection{Constraints with {\boldmath$f_{\rm duty}<1$}}\label{sec:results:fduty<1}

We now relax the assumption of continuous Eddington accretion ($f_{\rm duty}=1$). To save computational resources, we focus on the cut-power-law (CPL) spectrum, as we expect the main results to be qualitatively similar. To highlight both the uncertainties on accretion, and to illustrate the effect on the posteriors of less efficient growth, we consider specifically $f_{\rm duty} = (0.1, \, 0.3, \, 0.5)$ in Fig.~\ref{fig:fduty-full} (the posteriors for CPL with $f_{\rm duty} = 1$ can be seen in Fig.~\ref{fig:main-corner-cpl}, and are near-identical to $f_{\rm duty}=0.5$), with the corresponding parameter constraints summarized in Tab.~\ref{tab:cpl-results-fduty}. 
Perhaps the most stark consequence of a reduction in duty cycle is the effective disappearance of the negative $\tilde{f}_{\rm NL}$ mode, as can be seen in both corner plots. 

In order for the negative mode to source sufficiently abundant PBHs, a higher amplitude ($A$) is required, which in turn must be pushed to larger values of $k_p$ to avoid violating $\mu$-distortion constraints. When accretion is extremely efficient ($f_{\rm duty} \gtrsim 0.1$), the PBHs are capable of growing from a modest initial mass ($M_{\rm ini} \simeq \mathcal{O}(10^2 \, M_{\odot})$, corresponding to the earlier horizon re-entry times associated with a higher $k_p$), into the LRD-seeding black holes detected by JWST at $z_{\rm LRD} \simeq 7$. 
In contrast, once the duty cycle falls below roughly $10\%$, the black holes formed from the negative mode will fall short of the minimum mass requirement of $10^6 \,M_\odot$, and result in other astrophysical objects which may not appear in observations as LRDs. 
We note that perturbations that fail to meet the threshold to form PBHs may instead form not-quite-primordial black holes~\cite{Qin:2025ymc} or even small star clusters~\cite{Qin:2026jph} that may have other observable consequences on galactic dynamics or the CMB power spectrum~\cite{Weiner:2026fmw}, but consider a detailed study of these consequences to be out of scope for the present work.

For the positive $\tilde{f}_{\rm NL}$ mode, the contour sits at lower amplitudes for the primordial perturbation and is peaked at scales closer to the canonical region to which distortions are sensitive ($1 \, {\rm Mpc}^{-1} \lesssim k \lesssim 10^4 \, {\rm Mpc}^{-1}$). These lower $k_p$ values imply a heavier initial mass for the PBH population, so that even for $f_{\rm duty} = 0.1$ there is sufficient time to grow into SMBHs by the target redshift. Generically, the less efficient the accretion scenario, the more constraining CMB spectral distortions are for the primordially seeded SMBH scenario, as can be seen in the $\mu$ panels of Fig.~\ref{fig:fduty-full} where the posteriors push up against the \textit{COBE/FIRAS} bound as $f_{\rm duty}$ decreases. Towards even smaller values of the duty cycle parameter, we expect that there exists a threshold $f_{\rm duty} \leq f_{\rm duty}^{\textit{COBE}}$ below which the \textit{COBE/FIRAS} data itself already rules out supermassive PBHs. It is known, for example, that for (Gaussian) $\delta$-function enhancements of the primordial power spectrum with $f_{\rm duty} = 0$, \textit{COBE/FIRAS} provides a complete restriction on $M_{\rm PBH} > 10^4 \, M_{\odot}$ (see Fig.~11 of Ref.~\cite{Cyr2023}). Looking towards the future, while FOSSIL will provide some of the most sensitive and robust constraints on primordial SMBHs, other experiments will deliver interesting limits on an earlier timeline.
The BISOU \cite{Maffei:2021xur}, TMS \cite{TMS} and COSMO \cite{Masi:2021azs} experiments, primarily designed for detecting the sky-averaged Sunyaev-Zel'dovich $y$-distortion, will provide about an order of magnitude improvement over \textit{COBE/FIRAS}, placing the enhanced $\mu$-distortion coming from values of $f_{\rm duty} \simeq 0.1$ firmly within their sensitivity.

\begin{table*}
Constraints on a cut-power-law power spectrum with varying $f_{\rm duty}$
    \centering
    \begin{tabular} {|l|c|c|c|}
        \hline\hline 
        \rule{0pt}{1em} Parameter & $f_{\rm duty}=0.1$ & $f_{\rm duty}=0.3$ & $f_{\rm duty}=0.5$
        \\ \hline\hline
        \rule{0pt}{1em}
        {\boldmath$\log_{10} A$} 
        & $-2.76(-2.82)^{+0.09}_{-0.21}$ 
        & $-2.55(-2.79)^{+0.72}_{-0.54}$
        & $-2.54(-3.04)^{+0.71}_{-0.60}$
        \\
        \rule{0pt}{1em}
        {\boldmath$\log_{10}k_p$} 
        & $4.64(4.69)^{+0.08}_{-0.03}$ 
        & $5.01(5.34)^{+0.27}_{-0.17}$
        & $5.32(5.88)\pm{0.41}$
        \\
        \rule{0pt}{1em}
        {\boldmath$\tilde{f}_{\rm NL}$} 
        & $6.5(8.4)^{+2.9}_{-1.6}$ 
        & $3.8(6.5)^{3.6}_{-6.3}$
        & $3.3(9.6)^{+6.3}_{-6.5}$
        \\
        \hline
        \rule{0pt}{1em}
        $\tilde{f}_{\rm NL} \sqrt{A}$ 
        & $0.257(0.325)^{+0.066}_{-0.022}$ 
        & $0.10(0.26)^{+0.22}_{-0.04}$
        & $0.07(0.29)^{+0.25}_{-0.36}$
        \\
        \rule{0pt}{1em}
        $\log_{10}\mu$
        & $-4.47(-4.69)^{+0.07}_{-0.18}$ 
        & $-5.60(-6.92)^{+0.90}_{-0.54}$
        & $-6.5(-7.6)^{+1.4}_{-1.2}$
        \\
        \rule{0pt}{1em}
        $\log_{10}\mu_{\rm loop}$
        & $-5.64(-5.62)^{+0.10}_{-0.11}$ 
        & $-6.86(-7.64)^{+0.76}_{-0.74}$
        & $-7.8(-9.5)^{+1.4}_{-1.1}$
        \\
        \rule{0pt}{1em}
        $\log_{10}n_{\rm LRD}$
        & $-4.01(-4.00)^{+0.11}_{-0.09}$
        & $-4.00(-4.00)\pm 0.10$
        & $-4.00(-4.00)\pm 0.10$
        \\
        \hline
    \end{tabular}
    \caption{
    Marginalized parameter constraints and best-fit values on a cut-power-law power spectrum from \textit{COBE/FIRAS} $\mu$-distortion constraints in combination with JWST LRDs, for different assumptions on the accretion rates. Best-fit values are approximated by the best sample in the chain. For each parameter, the constraints are given as ``mean (best-fit) $\pm$ std.'' The parameters $k_p$ and $n_{\rm LRD}$ are given in units of (comoving) ${\rm Mpc}^{-1}$ and ${\rm Mpc}^{-3}$, respectively.
    }
    \label{tab:cpl-results-fduty}
\end{table*}

\section{\label{sec:level5}Discussions and Conclusions}

In this work, we have considered the possibility that the SMBHs present in the population of LRDs detected by JWST may have a primordial progenitor. 
Compared to a Pop III star origin, which would require the seeds to undergo sustained super-Eddington accretion in order to match the masses of the LRDs~\cite{2021ApJ...907L...1W,Bogdan:2023ilu}, PBHs have the advantage of being able to form at any mass and also having more time to accrete matter.
To assess whether this is a reasonable claim, we have performed a comprehensive suite of MCMC analyses for both lognormal (Eq.~\eqref{lognormalPR}) and cut-power-law (Eq.~\eqref{sharkPR}) enhancements to the primordial power spectrum, with the inclusion of local-type non-Gaussianity. Large enhancements to $\mathcal{P}_{\rm \mathcal{R}}$ are necessary to source an abundant population of heavy seed ($M \gtrsim 10^3 \, M_{\odot}$) PBHs, while at the same time it is known that these same enhancements inevitably give rise to significant CMB spectral distortions. 
The inclusion of non-Gaussianity allows for one to achieve the same number density of PBHs with a lower amplitude ($A$) power spectrum enhancement, which has the net effect of reducing the spectral distortion signal.

Crucially, it is only recently that tools have existed to allow for the robust determination of the $\mu$-type spectral distortion signal in the context of primordial non-Gaussianity. As discussed in Sec.~\ref{sec:level3.2}, the amplitude of the bispectrum contribution to the Silk-damping heating rate (which generates the spectral distortions) was shown in Ref.~\cite{Sharma:2024img} to be subdominant to the contribution coming from the power spectrum modification as given in Eq.~\eqref{eq:Rform}. This has provided confidence that the overall spectral distortion can be robustly computed as a sum of the ``tree" level contribution from $P_{\mathcal{R}_{\rm G}}$, and the ``loop" contribution from $P_{\mathcal{R}_{\rm G}^2}$. 

For a given power spectrum shape (lognormal or cut-power-law) and $f_{\rm NL}$, we simultaneously compute the expected $\mu$-distortion amplitude and the corresponding PBH mass function at formation. From here, we also include the effects of accretion onto the PBHs. For simplicity, we have assumed that the black holes grow at the Eddington limit between matter-radiation equality and $z=7$, and also consider the impact of a range of duty cycles in our results. PBHs which have grown beyond a mass of $M \geq 10^{6} \, M_{\odot}$ have sufficient Eddington luminosity to be detected by JWST, and thus would contribute to the LRD population. 

Recently, other works have appeared that also explore a primordial origin for a fraction of the high redshift SMBH population---our work is complementary to these. 
For example, some papers consider runaway mergers \cite{Ziparo:2024nwh} or sub-Eddington growth onto PBHs \cite{Hai-LongHuang:2024gtx, Dayal:2025aiv, Zhang:2025oyl} as a viable LRD production mechanism, but do not consider inevitable (large) spectral distortions that will be produced under generic PBH formation channels for their required mass functions. 
Other works \cite{Byrnes:2024vjt, Pritchard2025, Wang:2025hbw} defer discussions of accretion onto PBHs, even when their initial masses are $M\gtrsim 10^3 \, M_{\odot}$. 
Ref.~\cite{DeLuca:2025nao} discusses the formation of a single LRD (QSO1) through accretion and mergers of PBHs in a Gaussian setup, making use of Bondi accretion (which includes periods of super-Eddington growth). 
Finally, Ref.~\cite{Allegrini:2026jqt} considers the seeding of LRDs in a particular inflationary setup (the curvaton). This model is interesting in that it is possible to perform a full resummation of the non-Gaussianity introduced in the model, which adds to the robustness of their findings. 
All in all, we have performed the first MCMC of broad shape, non-Gaussian enhancements to the small-scale power spectrum, with the inclusion of Eddington-limited accretion, combining and adding to the relative strengths of these previous analyses into a powerful Bayesian framework. 

The conclusions of our analysis are in some ways dependent on the effective duty cycle of accretion after PBH formation. For efficient ($f_{\rm duty} \gtrsim 0.1$) accretion, the CPL spectrum gives rise to bimodal posteriors, where the two peaks correspond to positive and negative values of $f_{\rm NL}$. 
At these relatively large duty cycle parameters, PBHs can evade $\mu$-type distortion bounds in two ways. 
First, the large accretion rates permit smaller initial black hole masses (large values of $k_p$) and less (or no) non-Gaussianity. 
Secondly, cut-power-law enhancements to the spectrum which contain smaller $k_p$ values and more non-Gaussianity are also allowed, provided that the amplitude $A$ is slightly suppressed to keep the induced $\mu$ distortion at acceptable levels. 

When accretion becomes less efficient (e.g., for $f_{\rm duty} \lesssim 0.1$), one needs a mix of a larger overall $|f_{\rm NL}|$ and $A$, as well as a more red-shifted $k_p$, in order to produce an abundant population of PBHs. To remain consistent with $\mu$ constraints, $|f_{\rm NL}|$ is generically pushed to smaller values. As discussed earlier, however, small negative $f_{\rm NL}$ acts as a suppressor on the overall abundance (only large negative $f_{\rm NL}$ enhances the number density), and thus the negative lobe of the posteriors vanishes for small $f_{\rm duty}$. 
We expect that similar conclusions would hold for the case of lognormal enhancements.  

Put another way, as accretion becomes less efficient the bounds from CMB distortions become more constraining with respect to models that produce the same abundance of massive PBHs. This highlights the complementarity of spectral distortion and accretion bounds in constraining PBHs in this mass range. While accretion bounds have been derived for more sophisticated setups than ours, they rely on the assumption of an initially Gaussian distribution~\cite{Serpico:2020ehh, Serpico2024, Agius:2024ecw, Jangra:2024sif}. Recomputation of accretion bounds in these non-Gaussian setups is therefore a crucial future direction to robustly constrain primordial SMBHs.

In addition to determining whether PBHs that eventually seed LRDs can evade distortion constraints, we have also produced results considering only the $\mu$-type distortions for specifically non-Gaussian cut-power-law and lognormal enhancements to the primordial power spectrum on small scales. 
Regardless of whether or not one wishes to speculate on the origin of LRDs, these $\mu$-only posteriors provide valuable constraints for these specific enhancements. 

To conclude, we have performed the first set of MCMC analyses aimed at exploring the validity of (local type) non-Gaussian enhancements to the primordial power spectrum, specifically the cut-power-law and lognormal shapes. In this analysis, we have considered contributions to the $\mu$-type spectral distortion taking into account both tree level and one-loop contributions, and provided posteriors both with and without the requirement that the PBH population can seed the LRDs observed by JWST. We included a simple model of accretion, finding it to be a crucial ingredient for growing PBHs to $10^6 \, M_{\odot}$ by $z \simeq 7$. The distortion constraints depend on the duty cycle, and for $f_{\rm duty} \simeq 0.1$, the expected $\mu$ distortion to source the LRD population are very close to the current upper bound provided by the \textit{COBE/FIRAS} satellite. Thus, non-Gaussian supermassive PBHs present a highly interesting target for future spectral distortion experiments, such as TMS, BISOU, and FOSSIL, which will comprehensively probe this model over the coming years.

\section*{Acknowledgments}

This work made use of resources provided by subMIT at MIT Physics ~\cite{bendavid2025submitphysicsanalysisfacility}. 
SB was supported by the STFC under grant ST/X000753/1. 
BC was supported by an NSERC Banting Fellowship, as well as the Simons Foundation (Grant Number 929255). 
SRG is supported by an NSF Mathematical and Physical Sciences Ascend Fellowship  under Award No.~2317018. 
DLL was supported in part by a PhD Research Studentship from Research Manitoba, in part by a Canada Graduate Research Scholarship -- Doctoral from the Natural Sciences and Engineering Research Council of Canada (NSERC) [funding reference number CGRS D - 611992]. 
EM is supported in part by a Discovery Grant from the National Science and Engineering Research Council of Canada (NSERC). 
WQ is supported by the Simons Society of Fellows through Grant No. SFI-MPS-SFJ-00006250. 
Portions of this research were conducted in MIT's Center for Theoretical Physics -- a Leinweber Institute and supported in part by the U.S.~Department of Energy under Contract No.~DE-SC0012567.

\bibliography{bibliography}

\appendix

\section{\label{sec:levelA0}Computing the mass spectrum}

In this section, we provide the intermediate steps to go from the beta function in Eq.~\eqref{eq:beta_fun_comp} through to the mass spectrum in Eq.~\eqref{eq:mass_spectrum}. The bivariate Gaussian governing the statistics of small-scale perturbations in this setup is given explicitly by \cite{Iovino2024}
%
\begin{align}
    P_{\rm G}(\mathcal{C}_{\rm G}, \mathcal{R}_{\rm G}) &= \frac{{\rm e}^{-\mathcal{R}^2_{\rm G}/2\sigma_{\rm r}^2}}{2\pi \sigma_{\rm c} \sigma_{\rm r} \sqrt{1-\gamma_{\rm cr}^2}}\nonumber \\
    &\times {\rm exp}\left[- \frac{1}{2(1-\gamma_{\rm cr}^2)} \left( \frac{\mathcal{C}_{\rm G}}{\sigma_{\rm c}} - \frac{\gamma_{\rm cr} \mathcal{R}_{\rm G} }{\sigma_{\rm r}}\right)^2 \right].
\end{align}
%
This expression depends on the so-called correlators, $\sigma_{\rm c}^2 \equiv \langle \mathcal{C}_{\rm G} \mathcal{C}_{\rm G} \rangle$, $\sigma_{\rm r}^2 \equiv \langle \mathcal{R}_{\rm G} \mathcal{R}_{\rm G} \rangle$, and $\sigma_{\rm cr}^2 \equiv \langle \mathcal{C}_{\rm G} \mathcal{R}_{\rm G} \rangle$, given by
%
\begin{align}
\sigma_{\rm c}^2 &= \frac{4\Phi^2}{9} \int \frac{dk}{k} (kr)^4 W^2(k, r) T^2(k, r) \mathcal{P}_\mathcal{R}(k), \nonumber \\
\sigma_{\rm cr}^2 &= \frac{2\Phi}{3} \int \frac{dk}{k} (kr)^2 W(k, r) W_s(k, r) T^2(k, r) \mathcal{P}_\mathcal{R}(k), \nonumber \\
\sigma_{\rm r}^2 &= \int \frac{dk}{k} W_s^2(k, r) T^2(k, r) \mathcal{P}_\mathcal{R}(k). \nonumber 
\end{align}
%
Note that the correlators are where the precise shape (e.g. lognormal or cut-power-law) of the power spectrum becomes important in the compuation of the PBH spectrum. For completeness, the Fourier-space transfer and window functions are given by
%
\begin{align}
    T(k,\tau) &= 3 \left[ \frac{\sin(k\tau/\sqrt{3})-(k\tau/\sqrt{3}) \cos(k\tau/\sqrt{3})}{(k\tau/\sqrt{3})^3}\right], \nonumber\\
    W(k,r) &= 3 \left[ \frac{\sin(kr) - kr \cos(kr)}{(kr)^3}\right], \nonumber \\
    W_{\rm s}(k,r) &= \frac{\sin(kr)}{kr}. \label{eq:transfer-functions}
\end{align}
%
One thing to highlight is that the transfer function is typically computed at $T(k,\tau) = T(k,r_{m})$, where as discussed above, $r_{m}$ is the scale at which the compaction function itself is maximized. 

While it is important to give the full form of $P_{\rm G}$, its exact details are not necessary in order to derive Eq.~\eqref{eq:mass_spectrum}. The starting point for this is the usual relation between the beta function and the overall PBH energy density, given as
%
\begin{align}
    \Omega_{\rm PBH} &= \int \frac{{\rm d} M_{\rm H}}{M_{\rm H}} \left( \frac{M_{\rm eq}}{M_{\rm H}}\right)^{1/2} \beta_{\mathcal{C}} \nonumber \\
    &= \int \frac{{\rm d} M_{\rm H}}{M_{\rm H}} \left( \frac{M_{\rm eq}}{M_{\rm H}}\right)^{1/2} \nonumber \\
    &\hspace{5mm} \times \int_{\rm \mathcal{D}} {\rm d} \mathcal{C}_{\rm G} \, {\rm d} \mathcal{R}_{\rm G} \, P_{\rm G}(\mathcal{C}_{\rm G}, \mathcal{R}_{\rm G}) \, \frac{M_{\rm PBH}}{M_{\rm H}}.
\end{align}
%
The domain of integration $\mathcal{D}$ can be understood as requiring that two physical criteria be satisfied. First, one requires that the compaction exceeds a threshold. More specifically, 
%
\begin{align} \label{eq:bound_threshold}
    \mathcal{C} = \mathcal{C}_{\rm G} \frac{{\rm d} \mathcal{R}}{{\rm d} \mathcal{R}_{\rm G}} - \frac{\mathcal{C}_{\rm G}^2}{4\Phi}\left( \frac{{\rm d} \mathcal{R}}{{\rm d} \mathcal{R}_{\rm G}}\right)^2 > \mathcal{C_{\rm th}}.
\end{align}
%
The second criterion is that we wish to study type-I perturbations, in which the compaction monotonically increases towards its maximum value of $\mathcal{C} = \Phi$. As the relation between $\mathcal{C}$ and $\mathcal{C}_{l}$ is quadratic, this only occurs when
%
\begin{align} \label{eq:bound_branch}
    \mathcal{C}_l = \mathcal{C}_{\rm G} \frac{{\rm d} \mathcal{R}}{{\rm d} \mathcal{R}_{\rm G}} < 2\Phi.
\end{align}
%
Rearranging these bounds to provide the explicit integration domain
%
\begin{align}
    2 \Phi \left(\frac{{\rm d} \mathcal{R}}{{\rm d} \mathcal{R}_{\rm G}} \right)^{-1}\left(1- \sqrt{1- \frac{\mathcal{C}_{\rm th} }{\Phi}} \right) < &\mathcal{C}_{\rm G} <  2 \Phi \left(\frac{{\rm d} \mathcal{R}}{{\rm d} \mathcal{R}_{\rm G}} \right)^{-1}, \nonumber \\
    -\infty < &\mathcal{R}_{\rm G} < \infty .
\end{align}
%
With this in mind, we make a change of variable from $\mathcal{C}_{\rm G} \rightarrow M_{\rm PBH}$. (Recall that $M_{\rm PBH} = \mathcal{K} M_{\rm H} (\mathcal{C}-\mathcal{C}_{\rm th})^{\gamma}$.) The mapping is
%
\begin{align}
    \mathcal{C}_{\rm G} = &2\Phi \left(\frac{{\rm d} \mathcal{R}}{{\rm d} \mathcal{R}_{\rm G}} \right)^{-1} \nonumber\\
    &\times \left[1 - \sqrt{1- \frac{\mathcal{C}_{\rm th}}{\Phi} - \frac{1}{\Phi} \left(\frac{M_{\rm PBH}}{\mathcal{K} M_{\rm H}} \right)^{1/\gamma} }\right],
\end{align}
%
where the negative root is again taken to remain on the type-I branch. Performing the change of variables (which just amounts to inserting the proper Jacobian factor) yields a useful intermediate form: 
%
\begin{align}
    \Omega_{\rm PBH} &= \int \frac{{\rm d} M_{\rm H}}{M_{\rm H}} \left( \frac{M_{\rm eq}}{M_{\rm H}}\right)^{1/2} \int {\rm d} M_{\rm PBH} \, {\rm d}\mathcal{R}_{\rm G} \left(\frac{{\rm d} \mathcal{R}}{{\rm d} \mathcal{R}_{\rm G}} \right)^{-1} \nonumber \\
    &\times  \frac{1}{\gamma \mathcal{K} M_{\rm H}} \left(\frac{M_{\rm PBH}}{\mathcal{K} M_{\rm H}} \right)^{\frac{1-\gamma}{\gamma}} \left(\frac{M_{\rm PBH}}{M_{\rm H}} \right)\nonumber \\
    &\times \left( 1 - \frac{\mathcal{C}_{\rm th}}{\Phi} - \frac{1}{\Phi} \left( \frac{M_{\rm PBH}}{\mathcal{K} M_{\rm H}} \right)^{1/\gamma} \right)^{-1/2} P_{\rm G}(M_{\rm PBH},\mathcal{R}_{\rm G}).
\end{align}
%
With this change of variable, the limits on the $M_{\rm PBH}$ integral now become
%
\begin{align}
    0 < M_{\rm PBH} < \mathcal{K} M_{\rm H} (\Phi - \mathcal{C}_{\rm th})^{\gamma}. \nonumber
\end{align}
%
Finally, if we write ${\rm d} M_{\rm PBH} = M_{\rm PBH} \, {\rm d} \log M_{\rm PBH}$ and swap the order of integration between $M_{\rm H}$ and $M_{\rm PBH}$, we are left with
%
\begin{align}
    \Omega_{\rm PBH}= &\int {\rm d} \log M_{\rm PBH} \nonumber \\
    &\times \frac{\mathcal{K}}{\gamma} \int \frac{{\rm d} M_{\rm H}}{M_{\rm H}} \left( \frac{M_{\rm eq}}{M_{\rm H}}\right)^{1/2} \left(\frac{M_{\rm PBH}}{\mathcal{K} M_{\rm H}} \right)^{\frac{1+\gamma}{\gamma}} \nonumber \\
    &\times \left(1 - \frac{\mathcal{C}_{\rm th}}{\Phi} - \frac{1}{\Phi} \left[\frac{M_{\rm PBH}}{\mathcal{K} M_{\rm H}} \right]^{1/\gamma} \right)^{-1/2} \nonumber \\
    &\times \int {\rm d} \mathcal{R}_{\rm G} \left( \frac{{\rm d} \mathcal{R}}{{\rm d} \mathcal{R}_{\rm G}} \right)^{-1} P_{\rm G}(M_{\rm H}, \mathcal{R}_{\rm G}).
\end{align}
%
A trivial differentiation with respect to $\id \log M_{\rm PBH}$ yields Eq.~\eqref{eq:mass_spectrum}. With this new integration order, the $M_{\rm PBH}$ domain is passed to $M_{\rm H}$, which becomes
%
\begin{align}
    M_{\rm H} > \frac{M_{\rm PBH}}{\mathcal{K} (\Phi - \mathcal{C}_{\rm th})^{\gamma}} = M_{\rm H}^{\rm min}.
\end{align}
%

\section{Supplementary Plots}\label{app:supp_plots}

\begin{figure*}
    \centering
    \includegraphics[width=\linewidth]{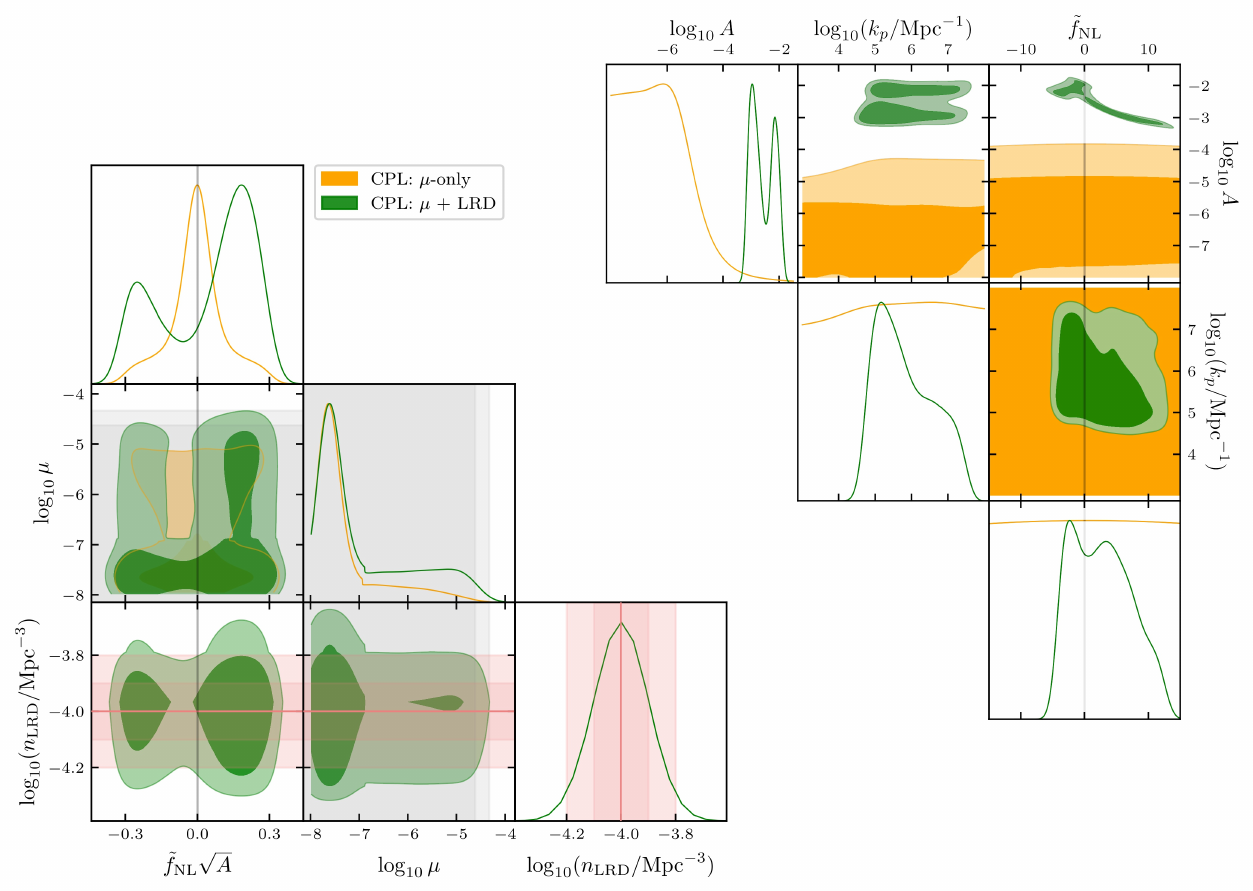}
    \caption{{\bf Cut-power-law model}: Constraints on the cut-power-law power spectrum with perturbative priors and $\mu$+LRD likelihoods. 
    The $1\sigma$ ($2\sigma$) constraints on $\mu$ and $n_{\rm LRD}$ are indicated by the light (dark) gray and light (dark) red shaded bands, respectively, with the red line representing the mean of the observed LRD abundance, $\log_{10}(n_{\rm LRD} / {\rm Mpc}^{-1})=-4.0$. The gray line highlights the Gaussian case ($\tilde{f}_{\rm NL}=0$).
    The bottom corner plot shows how the main observables ($\mu$ and $n_{\rm LRD}$) depend on the reduced model parameter $\tilde{f}_{\rm NL} \sqrt{A}$, and that they correctly fit observational constraints. The top corner plot shows the posteriors for the model parameters $A$, $k_p$, and $\tilde{f}_{\rm NL}$.}
    \label{fig:main-corner-cpl}
\end{figure*}

\begin{figure*}
    \centering
    \includegraphics[width=\linewidth]{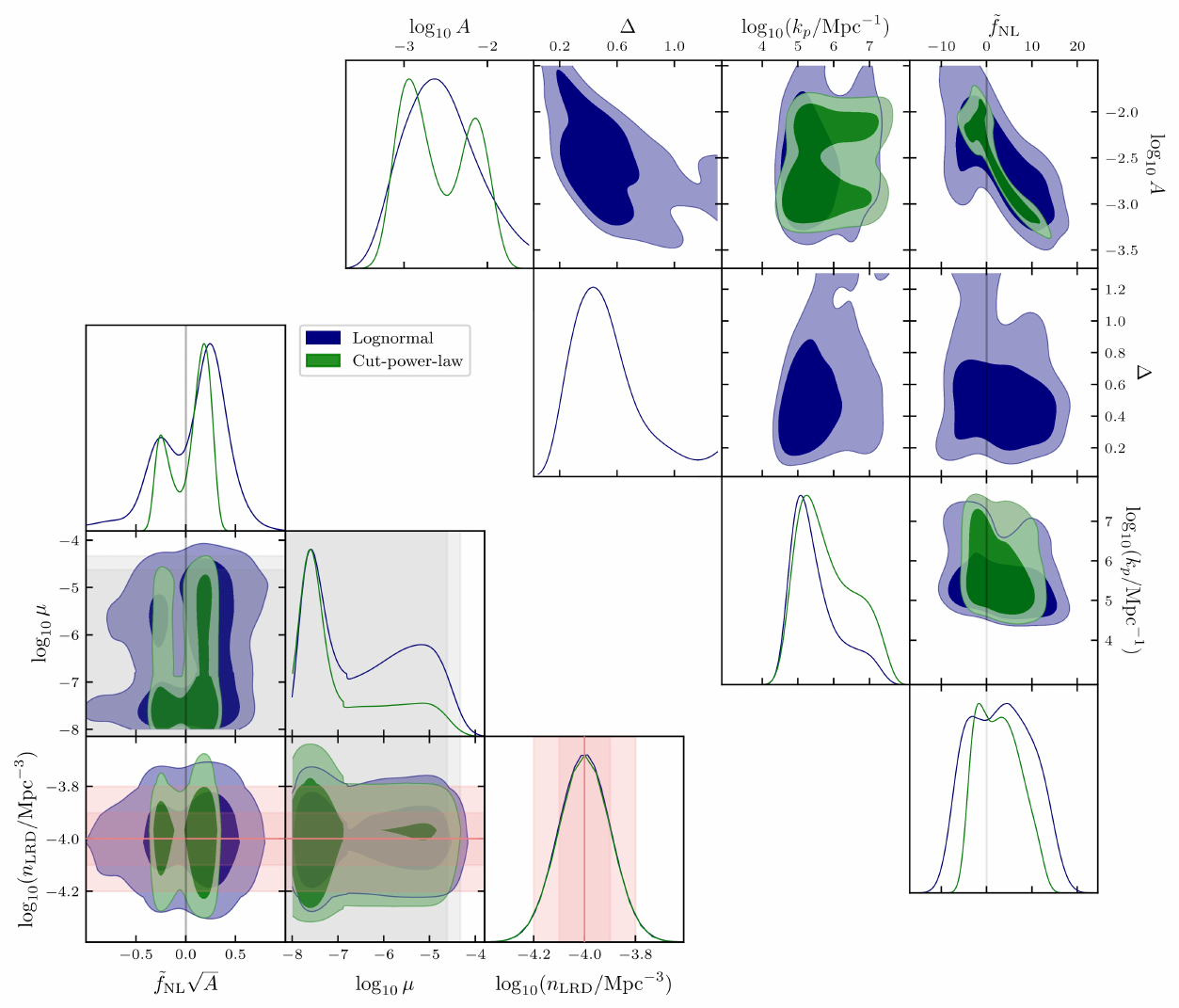}
    \caption{\textbf{Model Comparison:} Comparison of the posterior distributions from the MCMC for the lognormal versus cut-power-law power spectrum profiles with perturbative priors and $\mu$+LRD likelihoods. 
    The $1\sigma$ ($2\sigma$) constraints on $\mu$ and $n_{\rm LRD}$ are indicated by the light (dark) gray and light (dark) red shaded bands, respectively, with the red line representing the mean of the observed LRD abundance, $\log_{10}(n_{\rm LRD} / {\rm Mpc}^{-1})=-4.0$. The gray line highlights the Gaussian case ($\tilde{f}_{\rm NL}=0$).
    The bottom corner plot shows how the main observables ($\mu$ and $n_{\rm LRD}$) depend on the reduced model parameter $\tilde{f}_{\rm NL} \sqrt{A}$, and that they correctly fit observational constraints. The top corner plot shows the posteriors for the model parameters $A$, $\Delta$, $k_p$, and $\tilde{f}_{\rm NL}$.}
    \label{fig:model-comp}
\end{figure*}

\begin{figure*}
    \centering
    \includegraphics[width=\linewidth]{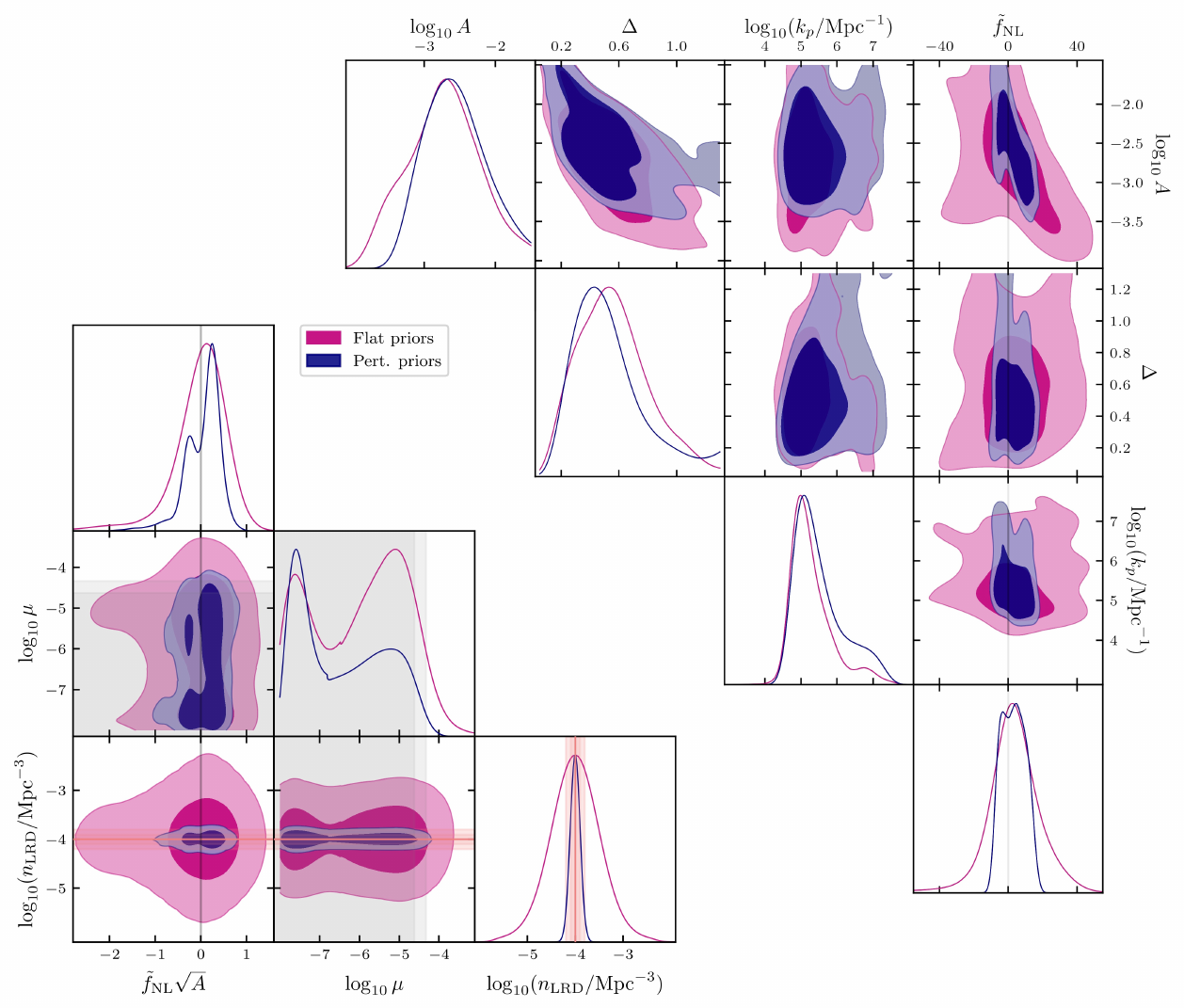}
    \caption{\textbf{Prior Comparison:} Comparison of the posterior distributions from the MCMC for the lognormal power spectrum profile with flat versus perturbative priors, with $\mu$+LRD likelihoods. 
    The $1\sigma$ ($2\sigma$) constraints on $\mu$ and $n_{\rm LRD}$ are indicated by the light (dark) gray and light (dark) red shaded bands, respectively, with the red line representing the mean of the observed LRD abundance, $\log_{10}(n_{\rm LRD} / {\rm Mpc}^{-1})=-4.0$. The gray line highlights the Gaussian case ($\tilde{f}_{\rm NL}=0$).
    The bottom corner plot shows how the main observables ($\mu$ and $n_{\rm LRD}$) depend on the reduced model parameter $\tilde{f}_{\rm NL} \sqrt{A}$, and that they correctly fit observational constraints. The top corner plot shows the posteriors for the model parameters $A$, $\Delta$, $k_p$, and $\tilde{f}_{\rm NL}$.}
    \label{fig:priors-comp}
\end{figure*}

\begin{figure}
    \centering
    \includegraphics[width=.55\linewidth]{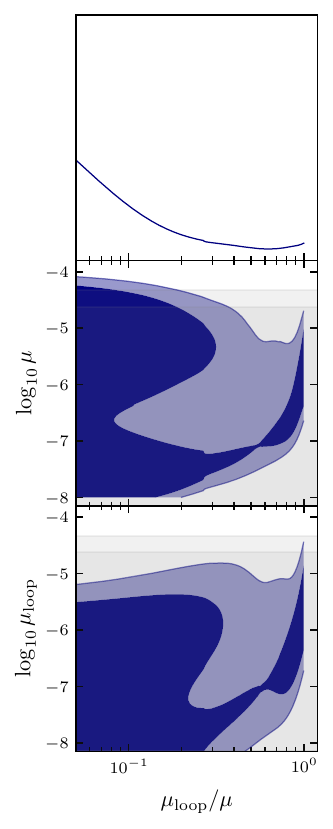}
    \caption{Posterior distribution for the ratio $\mu_{\rm loop}/\mu$, for the lognormal power spectrum profile with perturbative priors and $\mu$+LRD likelihood.
    The $1\sigma$ ($2\sigma$) constraints on $\mu$ are indicated by the light (dark) gray shaded bands.
    }
    \label{fig:mu}
\end{figure}

In the main text of this work, we highlight the results most relevant for understanding how non-Gaussian enhancements and accretion can produce PBH seeds consistent with $\mu$-type spectral distortions constraints and LRD abundances.
In this appendix, we include other useful results.
Fig.~\ref{fig:main-corner-cpl} shows the complete posteriors for the cut-power-law model.
Fig.~\ref{fig:model-comp} overlays the $\mu$+LRD posteriors for both the lognormal and cut-power-law enhancements for easy visual comparison.
Fig.~\ref{fig:priors-comp} demonstrates the effect of using uniform priors, compared to the perturbative priors described in the main text.
Fig.~\ref{fig:mu} shows the loop-contribution to $\mu$-type spectral distortions from lognormal enhancements and using the $\mu$+LRD likelihood.

\end{document}